\documentclass[twocolumn]{aastex62}
\usepackage{amsmath}
\usepackage{amssymb}
\usepackage{enumerate}   
\usepackage{booktabs}
\usepackage{xspace}
\usepackage{multirow}
\usepackage{color,soul}

\newcommand{\logg}{$\log g$\xspace}

\newcommand{\numax}{$\nu_{\text{max}}$\xspace}
\newcommand{\deltanu}{$\Delta \nu$\xspace}
\newcommand{\teff}{$T_{\text{eff}}$\xspace}
\newcommand{\gaia}{\textit{Gaia}\xspace}
\newcommand{\kepler}{\textit{Kepler}\xspace}
\newcommand{\tess}{\textit{TESS}\xspace}
\newcommand{\cannon}{\textit{The Cannon}\xspace}
\newcommand{\solmass}{M$_\odot$\xspace}
\newcommand{\solradius}{R$_\odot$\xspace}
\newcommand{\sollum}{L$_\odot$\xspace}

\newcommand{\swan}{\textit{The Swan}\xspace}
\newcommand{\sigmamad}{$\sigma_{\textrm{mad}}$\xspace}
\shorttitle{The Swan}
\shortauthors{Sayeed et al.}

\graphicspath{{./}{figures/}}

\begin{document}

% \title{Inference of Stellar Parameters using Data-Driven Modelling of Asteroseismic Power Spectra}
% \title{Inference of stellar surface gravity from granulation signal for 20,437 Kepler dwarfs and giants} %using linear regression?
\title{The Swan: Data-driven inference of stellar surface gravities for cool stars from photometric light curves}

%% Note that the corresponding author command and emails has to come
%% before everything else. Also place all the emails in the \email
%% command instead of using multiple \email calls.
\correspondingauthor{Maryum Sayeed}
\email{maryum.sayeed@alumni.ubc.ca}

\author[0000-0001-6180-8482]{Maryum Sayeed}
\affiliation{Department of Physics and Astronomy, University of British Columbia, 6224 Agricultural Road, Vancouver, BC V6T 1Z1, Canada}

\author[0000-0001-8832-4488]{Daniel Huber}
\affiliation{Institute for Astronomy, University of Hawai`i, 2680 Woodlawn Drive, Honolulu, HI 96822, USA}

\author[0000-0001-7339-5136]{Adam Wheeler}
\affiliation{Department of Astronomy, Columbia University, 550 West 120th Street, New York, NY 10027, USA}

\author[0000-0001-5082-6693]{Melissa Ness}
\affiliation{Department of Astronomy, Columbia University, Pupin Physics Laboratories, New York, NY 10027, USA}
\affiliation{Center for Computational Astrophysics, Flatiron Institute, 162 Fifth Avenue, New York, NY 10010, USA}

%% Note that RNAAS manuscripts DO NOT have abstracts.
\begin{abstract}

Stellar light curves are well known to encode physical stellar properties. Precise, automated and computationally inexpensive methods to derive physical parameters from light curves are needed to cope with the large influx of these data from space-based missions such as \kepler and \textit{TESS}. Here we present a new methodology which we call \swan, a fast, generalizable and effective approach for deriving stellar surface gravity (\logg) for main sequence, subgiant and red giant stars from \kepler light curves using local linear regression on the full frequency content of \kepler long cadence power spectra. With this inexpensive data-driven approach, we recover \logg to a precision of $\sim$0.02 dex for 13,822 stars with seismic \logg values between 0.2-4.4 dex, and $\sim$0.11 dex for 4,646 stars with \gaia derived \logg values between 2.3-4.6 dex. We further develop a signal-to-noise metric and find that granulation is difficult to detect in many cool main sequence stars (\teff $\lesssim$ 5500 K), in particular K dwarfs. By combining our \logg measurements with \gaia radii, we derive empirical masses for 4,646 subgiant and main sequence stars with a median precision of $\sim$7\%. Finally, we demonstrate that our method can be used to recover \logg to a similar mean absolute deviation precision for \tess-baseline of 27 days. Our methodology can be readily applied to photometric time-series observations to infer stellar surface gravities to high precision across evolutionary states.

\end{abstract}
%% See the online documentation for the full list of available subject keywords and the rules for their use.

% \tableofcontents

\section{Introduction} 
Over the last decade, there has been a substantial increase in photometric time-series observations from ground-based and space-based telescopes. With past and existing missions like CoRoT \citep{corot_2006, corot}, \kepler \citep{kepler_borucki, kepler_koch}, TESS \citep{tess}, and the upcoming PLATO \citep{plato}, we now have time-series data available for hundreds of thousands of stars. Such an abundance of time-domain data provides valuable insight about stellar physics and interiors, and advances our knowledge of stellar properties and distribution in the galaxy. Furthermore, understanding properties of host stars has important implications for exoplanet studies \citep{johnson_2010, huber_2013, fulton_2018}.

One of the most successful methods to determine stellar properties from time-series data is asteroseismology, the study of stellar oscillations \citep{brown_1994, astero_book_2010, chaplin_2013, hekker_2017}. Using high-cadence photometry, two characteristic observables can be measured from the power spectrum: the frequency of maximum oscillation power, \numax, and large frequency separation between two modes of the same spherical degree, \deltanu \citep{brown_1991, belk_2011, ulrich}. These two quantities allow us to calculate stellar masses and radii through scaling relations \citep{kjeldsen_bedding, stello_2009, kallinger_2010, huber_2011, casagrande_2014, mathur_2016, anders_2017}. For instance, \cite{kallinger_2010} used $\sim$137 days of \kepler data to derive asteroseismic mass and radius for red giants to within 7\% and 3\%, respectively; \citet{silva_2017} used individual oscillation frequencies to determine mass, radius and age for 66 main sequence stars in the \kepler sample to an average precision of $\sim$4\%, $\sim$2\% and $\sim$10\%, respectively; \cite{huber_2013} used asteroseismology to determine fundamental properties of 66 \kepler planet-candidate host stars, with uncertainties of $\sim$3\% in radius and $\sim$7\% in mass. 

Oscillation amplitudes increase with stellar luminosity, thus making it easier to measure oscillatons in red giants \citep{kjeldsen_bedding, huber_2011}. However, using similar methods to calculate mass, radius and age for dwarfs becomes difficult due to their low luminosity. Moreover, oscillations in dwarfs occur at higher frequencies which cannot be detected in long cadence \kepler data. Short cadence data, while extremely useful for dwarfs, are available for only a small number of stars. Therefore, a sample of asteroseismic parameters for dwarfs is severely limited.  

Alternatively, we can use granulation to derive stellar properties. The granulation signal -- often referred to as the ``noise" term or ``background'' in the power spectrum -- is caused by convection on the stellar surface. Granulation signal correlates strongly with stellar surface gravity (\logg), and can be used to calculate \logg from asteroseismic signals \citep{mathur_2011, kjeldsen_2011}. This method was used to determine fundamental parameters of dwarfs in \cite{bastien_2013}, but was limited by RMS measurements in the time-domain at a fixed frequency. \cite{kallinger_2016} further developed the technique to estimate \logg from the auto-correlation function (ACF), but their analysis was limited to short cadence \kepler data of main sequence stars. \cite{pande_2018} transitioned to frequency space to calculate \logg from the granulation signal, but used a limited region of the power spectrum in their method. \cite{bugnet_2018} also estimated surface gravities from global power in the spectrum using a random forest regressor, but with a small bias for higher luminosity and main sequence stars. While largely successful, most current methods that use granulation to extract fundamental stellar parameters do not use the full information contained in the power spectrum, are limited to certain evolutionary states, or do not quantify whether granulation has been detected. 

Recent work in data-driven modelling techniques, such as \cannon, has improved the accuracy and efficiency to which we derive stellar parameters from large astronomical data sets. \cannon is a data-driven approach which initially has been developed to infer \logg, mass, effective temperature (\teff) and chemical abundances from spectroscopic data \citep{ho_2017a, ho_2017b, casey_2016, casey_2017, wheeler_2020a, tawny_2020, rice_2020}. \cite{ness_2018} presented the first application of \cannon on \kepler time-series data, and modeled the auto-correlation function to recover \logg to a precision of $<0.1$ dex. However, their analysis was applied to red giants only. 

In this paper, we expand on the work by \cite{ness_2018} to estimate \logg from the power spectrum of main sequence, subgiant and red giant stars using long cadence \kepler data. We develop a new method which we call \swan \footnote{inspired by Henrietta Swan Leavitt, a pioneer of variable star astronomy using photometry}, which employs local linear regression to infer \logg using a localised training set around each star under consideration, and derive stellar masses using radii from \gaia. We compare the success of previous data-driven methods like \cannon to our model, and investigate the effect of photometry baseline on \logg inference. 

\section{Observations}
\label{sec:data}
\subsection{Target Selection}
To use data-driven inference techniques, we create a reference set of main sequence, subgiant and red giant stars for which \logg is well known and granulation can be detected. We select 16,947 stars with asteroseismically derived stellar surface gravities in \cite{mathur_2017}, who compiled seismically derived stellar parameters from various catalogs \citep{bruntt_2012, thygesen_2012, stello_2013, chaplin_2014, huber_2011, huber_2013, huber_2014, pinsonneault_2014, casagrande_2014, silva_aguirre_2015, mathur_2016}, and 16,094 stars with seismically derived masses, radii, and surface gravities in \citet{yu_2018}. By cross-referencing both catalogs, we select 19,273 stars with asteroseismic surface gravities. This sample is hereafter referred to as the seismic sample. 

Furthermore, low frequency stellar variability that is not due to granulation can mask the granulation signal that we want to measure. To measure only the granulation background, we removed any rotating stars \citep{mcquillan_2013}, classical pulsators \citep{debosscher_2011, murphy_2019}, eclipsing binaries \citep{abdul_2016} and \kepler exoplanet host stars \citep{kepler_hosts}. In total, we removed 1,851 stars exhibiting non-granulation like variability from the seismic sample. To ensure we use the most recent stellar parameters, we removed 998 stars not found in \citet{berger_2018}, 216 stars with photometry less than 89 days and 1,788 stars for which we could not find Quarter 7-11 data. We also removed 252 stars that performed below a 50\% duty cycle (see Section \ref{sec:data_prep}). The final seismic sample contains 14,168 stars. 

\begin{figure}[t!]
\begin{center}
\includegraphics[width=0.5\textwidth,angle=0]{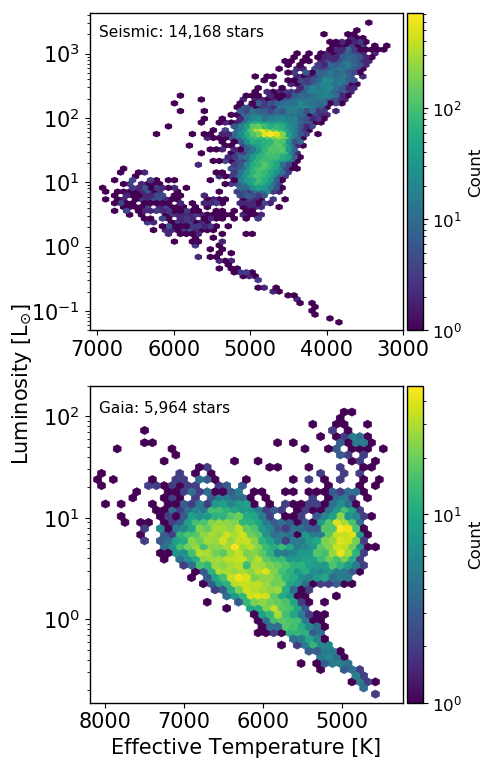}
\caption{Hertzprung-Russell (H-R) diagrams of stars in our reference sample. \textit{Top}: Luminosity versus effective temperature for 14,168 in the seismic sample. \textit{Bottom}: Luminosity versus effective temperature for 5,964 stars in the \gaia sample. Colour-coding represents number density.}
\label{hr_both}
\end{center}
\end{figure}

In the top panel of Figure \ref{hr_both}, we plot the distribution of stars in the seismic sample on a Hertzsprung-Russell (H-R) diagram. Stellar \logg values for this sample are calculated using asteroseismic scaling relations. This sample contains mostly giants due to their high signal-to-noise ratio which makes it possible to measure global seismic observables, such as \numax and \deltanu, to derive stellar properties.

To expand the sample of dwarfs and subgiants, we chose stars for which granulation has been successfully detected by \citet{pande_2018}, hereafter referred to as the \gaia sample. The catalog contains 15,109 stars, but 6,736 stars overlap with the seismic sample. Because stellar parameters measured from asteroseismology are more accurate and precise, we classified overlapping stars as an asteroseismic measurement due to their high precision. Out of the remaining 8,373 stars, we removed 1,834 stars that exhibit other types of variability. We further removed 392 stars not found in \citet{berger_2018} to ensure we use recently derived stellar parameters. Out of the remaining 6,147 stars, we removed 84 stars with photometry less than 89 days, 88 stars performing below a 50\% duty cycle (see Section \ref{sec:data_prep}), 10 stars for which we could not retrieve Quarter 7-11 data, and 2 stars for which there was no \logg value available in \citet{berger_2020}. The final \gaia sample contains 5,964 stars. 

The bottom panel of Figure \ref{hr_both} shows the distribution of stars in the \gaia sample. Compared to the seismic sample, the \gaia sample contains many more dwarfs since it does not suffer from the seismic detection bias \citep{chaplin_2011}. The \logg values for this sample are calculated from isochrone fitting in \cite{berger_2020} who used recent MESA Isochrones and Stellar Tracks (MIST) models \citep{choi_2016, dotter_2016, paxton_2011, paxton_2013, paxton_2015}, SDSS g and 2MASS $K_s$ photometry, \gaia DR2 parallaxes, red giant evolutionary state flags and spectroscopic metallicities to derive a homogeneous set of stellar parameters. Median fractional uncertainties for derived stellar surface gravity are on the order of 0.05 dex, with \logg distribution peaking at 4.24 dex. 

In this work, giants are defined as stars with a \logg below 3.5 dex, while dwarfs and subgiants have a \logg above or equal to 3.5 dex. We obtained \logg values for most dwarfs from \citet{berger_2020}, and \logg values for subgiants and giants from \citet{mathur_2017} and \citet{yu_2018}. If a star was available in all three catalogs, asteroseismic measurements were prioritized. If the \logg was unavailable in the seismic catalogs, the label was obtained from \citet{berger_2020}. Our final reference sample contains 20,132 stars exhibiting granulation with \logg measurements from asteroseismology \citep{mathur_2017, yu_2018} and isochrone fitting \citep{berger_2020}. 

\begin{figure}[t]
\begin{center}
\includegraphics[width=0.5\textwidth]{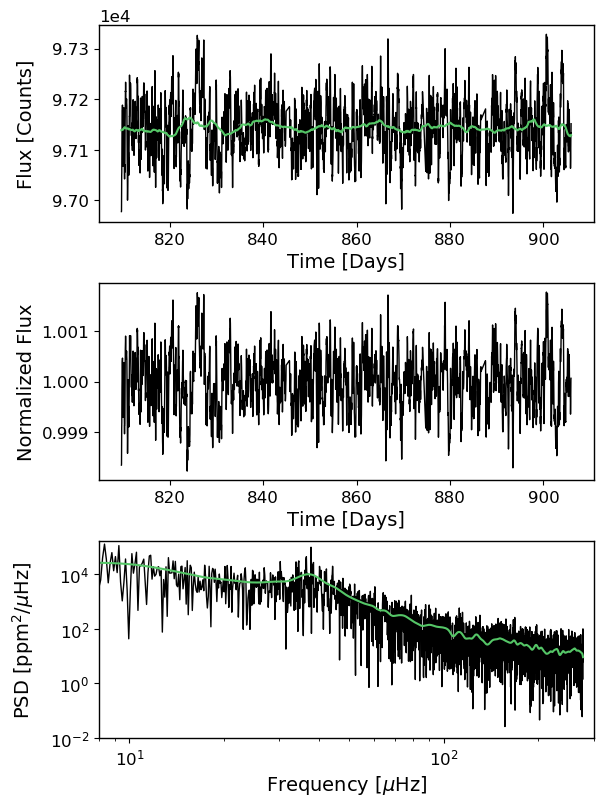}
\caption{Overview of steps taken to prepare the data. \textit{Top}: Raw \kepler long cadence light curve of \kepler ID 8812919. Black curve is the outlier corrected light curve and the green line is the smoothed light curve using a width of $\sim$8 days. \textit{Middle}: Normalized flux after removing long-periodic (low-frequency) variations. \textit{Bottom}: Power spectrum of normalized light curve in black. The green line represents the power spectrum smoothed by 2 $\mu$Hz and effectively traces the granulation signal of the star.}
\label{data_prep}
\end{center}
\end{figure}

\begin{deluxetable}{cc}[t!]
\tabletypesize{\footnotesize}
    \tablecolumns{10}
    \tablewidth{0pt}
    \tablecaption{\label{tb:whitenoise_vs_kp} White noise level as a function of \kepler magnitude calculated using 3,270 \kepler stars observed in short cadence.}
    \tablehead{\colhead{Kepler Magnitude} & \colhead{White Noise Level [ppm$^2/\mu$Hz]}}
    \startdata
    8.0 & 0.17 \\
    8.1 & 0.19 \\
    8.2 & 0.21 \\
    ... & ... \\
    15.8 & 963.30 \\
    15.9 & 1109.85 \\
    16.0 & 1278.83 \\
    \enddata
    \tablecomments{The full table in machine-readable format can be found online.}
\end{deluxetable}

\begin{figure*}[t!]
\begin{center}
\includegraphics[width=1\textwidth,angle=0]{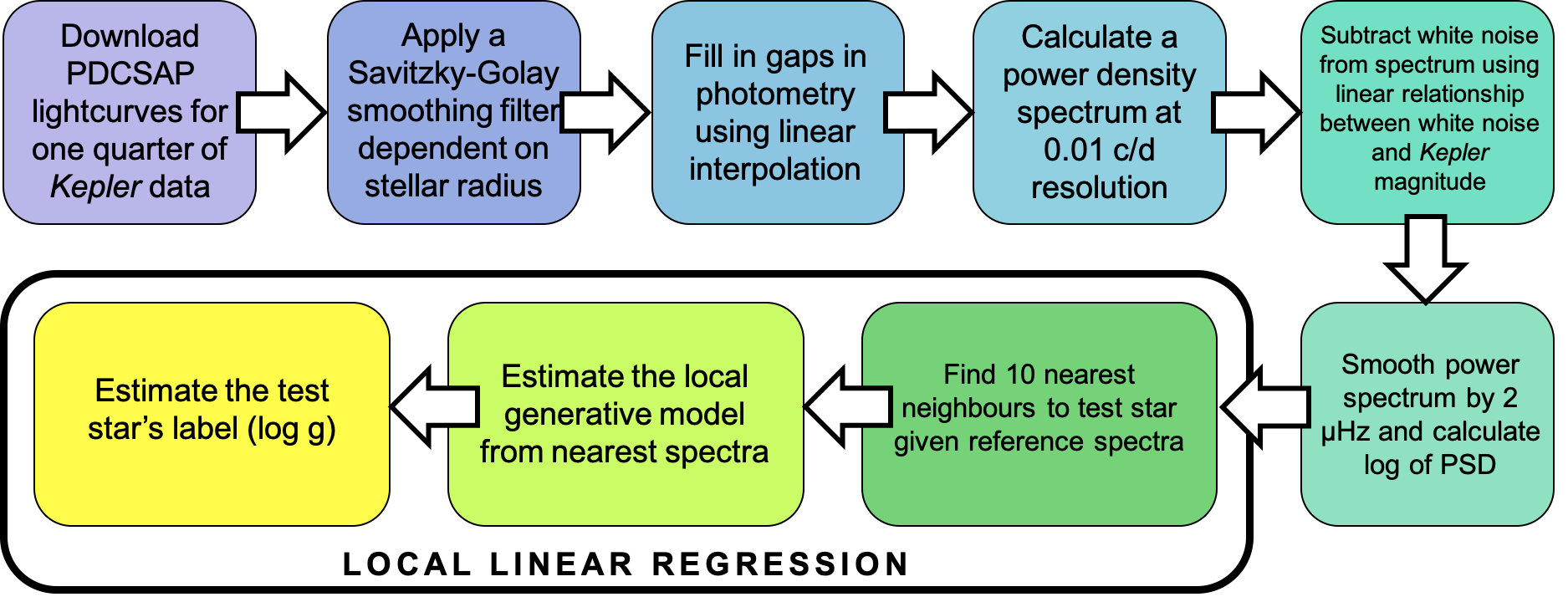}
\caption{Overview of \swan pipeline including details on the preparation of input data and application of local linear regression as described in Sections \ref{data_prep} and \ref{sec:llm_lc}.}
\label{pipeline}
\end{center}
\end{figure*}

\subsection{Data Preparation}
\label{sec:data_prep}
To predict stellar surface gravity from power spectra, we prepared long cadence \kepler data (29.4 minute sampling) to train and test our model. We use a single quarter of \kepler data, or $\sim$90 days, to minimize data-processing and validate our model on the shortest available photometry baseline. Figure \ref{data_prep} shows the steps we followed for data preparation as described below:

\begin{enumerate}[(i)]
    \setlength\itemsep{0em}
    \setlength\parskip{0em}
    \item We downloaded Pre-Search Data Conditioning Simple Aperture Photometry (PDCSAP) from MAST for stars with Quarter 9 data. We only selected stars for which we had \textit{Gaia} stellar parameters as found in \citet{berger_2018}. For stars without Quarter 9 data, we tried to find data in adjacent quarters and removed any stars from our sample for which there was no Quarter 7-11 data available. For the seismic sample, the number of stars in Quarter 9-11 were 10,104, 3,808 and 256, respectively. For the \gaia sample, the number of stars in Quarter 9-11 were 5,622, 338 and 4, respectively. 
    \item We only kept data points with good quality flags (flags = 0).
    \item We removed any stars that had a duty-cycle less than 50\%.
    \item We performed a running sigma-clipping routine with 50 data points in each subset and removed outliers greater than 3$\sigma$. 
    \item We applied a smoothing Savitzky-Golay filter to remove any long-periodic (low-frequency variations). To create a relation between smoothing filter and stellar radius, we used the radius and \numax measurements for stars in \citet{yu_2018}, along with the empirically derived smoothing function d=0.61+0.04\numax as mentioned in \citet{yu_2018}. The smoothing width for each star varied according to the following relation given a stellar radius in solar units: 
    \begin{equation}
    d \; [\mu \textrm{Hz}] = 0.92+23.03 \; e^{-0.27 \;\textrm{R}/\textrm{R}_\odot}
    \end{equation}
    The filters ranged from 0.5 - 12.6 days, where a filter of $\sim$12.6 days was applied to stars with stellar radius above $\sim$30 \solradius. 
    \item Any remaining gaps in the data created by bad quality flags were filled in using linear interpolation to avoid leakage of power towards higher frequencies in the power spectrum \citep{garcia_2014, pires_2015}. We followed the interpolation by removing any remaining data points in the smoothed and normalized light curve greater than 5$\sigma$ from the flux median.
    \item We calculated a power density spectrum with a frequency spacing of $\sim$0.12 $\mu$Hz (or $\sim$0.01 c/d) corresponding to a critically sampled spectrum. 
    \item We subtracted the contribution from white noise using a linear relationship between \kepler magnitude (Kp) and white noise (see Table \ref{tb:whitenoise_vs_kp}) where Kp values were obtained from \kepler Input Catalog (KIC, \citet{brown_2011}. The relation was created using one quarter of \kepler data from 3,270 stars observed in short cadence and prepared using steps i-vii. The white noise was calculated as the average power between frequencies 6,000-8,000 $\mu$Hz. We subtracted the white noise from the unsmoothed spectrum for frequency bins with granulation power greater than the white noise. This step ensured that we recovered most of the granulation signal in the spectrum especially for faint stars.
    \item We smoothed the power spectrum by 2 $\mu$Hz to remove sharp peaks in amplitude. We experimented with the smoothing value, and chose a value that ensured we did not lose information in the spectrum, that is, did not compromise the precision of our inferred labels (as parameterized by the RMS scatter and median standard deviation). 
\end{enumerate}

\section{Local Linear Regression applied to Light Curves}
\subsection{\swan}
\label{sec:llm_lc}
Local models (sometimes also referred to as ``kernel methods''; see e.g. \citealp{Hastie_Tibshirani_Friedman_2009} chapter 6 for an introduction) capture a non-linear relationship between labels and data by combining linear models to the small region around each datapoint.  Local Linear Models (LLM) find a natural application in large data sets where the data are a globally complex, but smooth function of labels.  By ``zooming in'' to a regime where the model is approximately linear, local linear regression can capture globally non-linear behavior with a mathematically simple model (see, e.g. \citealp{DH2010} for a review of linear regression).   While local methods often fail in application to high dimensional data, the smoothed stellar power spectra of time-domain photometry (light curves) have few degrees of freedom \citep{anderson_1990}. For our application, the data is the noise-corrected power spectrum of the light curves and the label we infer is the \logg parameter for each. 

In brief, for each test object, \swan uses training data in the local neighbourhood to infer labels via regression.
Given $\mathbf{G}$, taken to be a vector of the points in the log power spectrum of a star for which we would like to obtain labels, we find the $k$ power spectra in the training set with the smallest Euclidean distances, $d$, from $\mathbf{G}$:
\begin{equation} \label{eq:dist}
    d = |\mathbf{G} - \mathbf{G}_\text{train}| \quad ,
\end{equation}
where $G_\text{train}$ is the Power Spectral Density (PSD) vector from the training set.
We take $\Gamma$ to be the matrix whose rows are the neighboring training spectra, and $L$ to be the matrix whose rows are the label vectors of the neighbors. Since we use a single label, $L$ is a vector of the \logg values for each neighboring star.
We standardize $\Gamma$ and $\mathrm{G}$ by re-expressing them in coordinates such that the mean neighboring spectrum is zero, and the standard deviation of the power spectral density among the neighbors is one at each frequency.
Our prediction for the label vector (or label) associated with $\mathbf{G}$ is then given by,
\begin{equation} \label{eq:model}
    \widehat\ell = (L^+ \Gamma)^+ \mathbf{G} \quad ,
\end{equation}
where ``$^+$'' denotes the Moore-Penrose pseudoinverse defined for real orthogonal matrices as 
\begin{equation}
    A^+ = (A^T A)^{-1} A^T \quad .\footnote{As always, explicit numerical matrix inversion should be avoided when possible.  A \texttt{solve} function should be used instead.}
\end{equation}

This linear operation has a simple interpretation.
$L^+\Gamma$ is the maximum-likelihood estimate (MLE) for the matrix that maps labels (values of \logg) to power spectra, assuming that the labels are known exactly and the PSD at each frequency in each power spectrum has independent equal Gaussian uncertainty.
While the measurement uncertainty on each power spectrum is neither Gaussian nor homoscedastic, it is much smaller than the model prediction error, indicating that a more careful treatment would be unhelpful.
$\widehat\ell$ is the MLE for the labels associate with $\mathbf{G}$ given $L^+\Gamma$; it is the label(s) that yields a predicted power spectrum closest to $\mathbf{G}$.
The best-fit model spectrum (that generated exactly from $\widehat\ell$) is given by $\widehat\ell^T (L^+\Gamma)$.
We tried using a discriminative linear model ($\widehat\ell = \mathbf{G}^T \Gamma^+ L$) rather than a generative one, but found that it performed slightly worse in cross-validation.

In Figure \ref{pipeline}, we show a schematic of \textit{The Swan}, including the steps taken to prepare the data for local linear regression. To summarize, the input to \swan is the logarithmic power spectral density and labels that describe the data.  To estimate the \logg from a power spectrum, \swan finds the most similar spectra in the training set (those which minimize Equation \ref{eq:dist}).
It then performs linear regression \emph{on those spectra only} to infer the relationship between \logg and power spectral density, and uses that generative model to estimate the \logg of the spectrum at hand (Equation \ref{eq:model}).

We experimented with the number of nearest neighbours (ie. $k$ = 5, 10, 20, 50 and 100) and found there was only a slight difference in \swan's performance for small $k$ compared to large $k$, and worsened for $k$ above 50. We chose $k = 10$ as the number of nearest neighbours. This value of $k$ is computationally trivial, and ensures there are enough stars to model well the relationship between the power spectral density and the \logg label.

\subsection{Seismic sample}
\label{sec:seismic_results}

\begin{figure}[t!]
\begin{center}
\includegraphics[width=0.5\textwidth,angle=0]{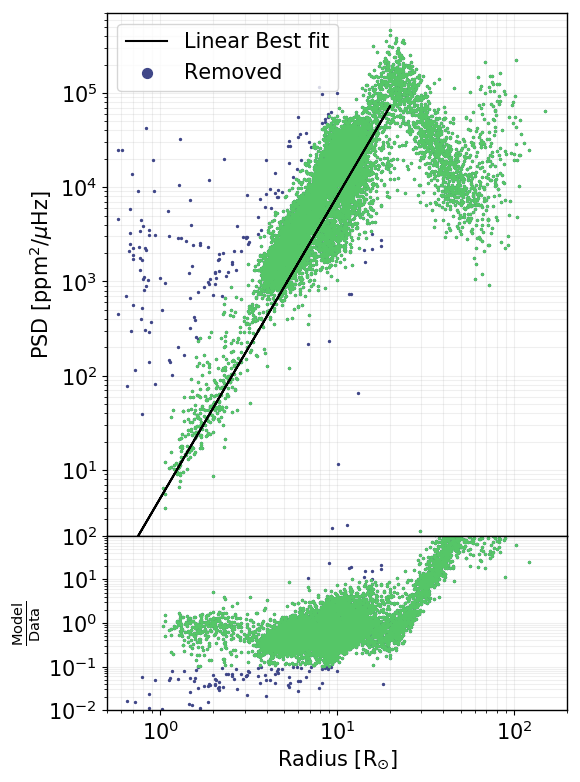}
\caption{Mean power spectral density between 10-12 $\mu$Hz as a function of stellar radius for stars in the seismic sample. Stars dominated by variability other than granulation are identified as outliers which are shown as blue data points, while green data points represent stars that pass our test for no additional variability (and below 20 \solradius represent stars within 10 ppm$^2$/$\mu$Hz of the best fit line).}
\label{astero_pr}
\end{center}
\end{figure}

\begin{figure*}[t!]
\begin{center}
\includegraphics[width=1\textwidth,angle=0]{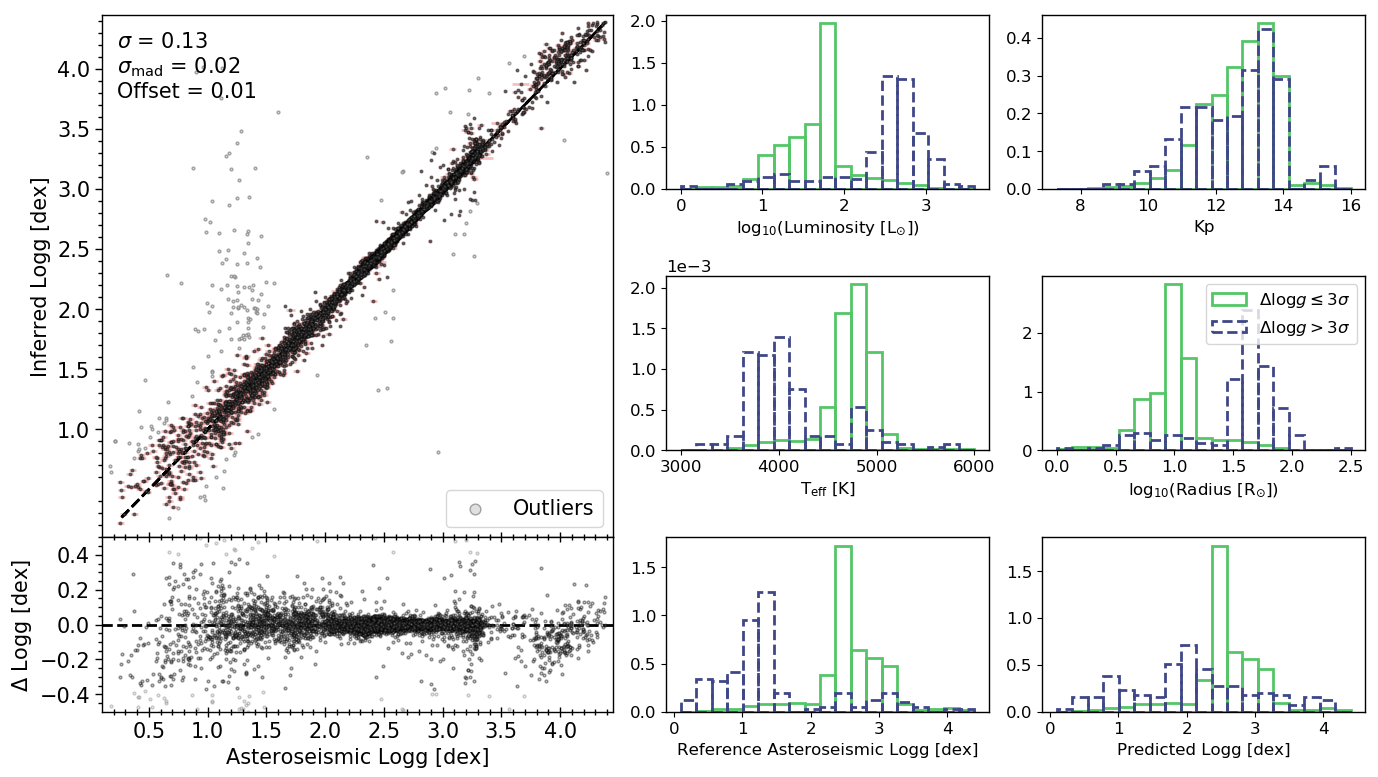}
\caption{\textit{Left}: Surface gravities for the seismic sample inferred with \swan, compared to the asteroseismic \logg reference labels. The top left hand corner shows the precision and bias metrics of the inference, where $\sigma$ is the root mean square (RMS) and \sigmamad is the robust standard deviation for 14,003 stars (including outliers). Outliers, defined as stars with a difference between their reference and inferred \logg above $3\sigma=0.38$ dex, are shown in grey and errors in asteroseismic \logg are shown in red. The bottom left panel shows the difference between the reference asteroseismic \logg and inferred \logg. \textit{Right}: Histograms demonstrating the properties of stars in the seismic sample. The green solid bars show normalized counts for non-outliers and the purple dashed bars show normalized counts of outliers.}
\label{ast_result}
\end{center}
\end{figure*}

To validate our model, we first applied \swan on the seismic sample where \logg is measured using asteroseismic scaling relations \citep{mathur_2017, yu_2018}. These reference \logg values serve to label each test star, by building a model using its nearest neighbours in data space as described in Section \ref{sec:llm_lc}. This is equivalent to cross-validation, since a test star itself is not used to estimate its own label. We can then compare the reference \logg value for the test star to our inferred \logg value. 

In generating results of the seismic analysis, we made some additional quality cuts on the data beyond the complete reference set. While we had first removed stars from our sample that showed non-granulation like variability and also applied a high-pass filter to attenuate the rotation signal, our initial cross-correlation tests of our inferred versus reference \logg label showed a number of outliers, with high amplitude at low frequencies in their power spectra. These stars correspond to fast rotators with prevalent harmonics in the spectrum. This makes them sufficiently different from the rest of the data since the linear model can not capture the power amplitude as a function of the \logg label. To screen for stars with harmonics in the spectrum, we asserted a linear relationship in log-log space between the average power spectral density at low frequencies and stellar radius from \gaia, and modeled it using the following relationship:
\begin{equation}
    \log_{10} (\text{PSD}/(\text{ppm}^2/\mu \text{Hz}))=3.2 \log_{10} (\text{R/R}_{\odot}) + 0.7
\label{eq:pr_eq}
\end{equation}

Figure \ref{astero_pr} shows mean power spectral density between 10-12 $\mu$Hz as a function of stellar radius. The turnoff at 20 \solradius corresponds to the impact on the data of applying this filter on stars with large radii. The best fit line was found by fitting the relationship between PSD and stellar radius for radii below 20 \solradius. We removed any stars with a power spectral density larger than 10 $\text{ppm}^2/\mu \text{Hz}$ from the best fit line for radii below 20 \solradius, and kept all stars above this threshold. Through this filter, we removed $\sim$1.1\% of the seismic sample, or 160 from an original 14,168 stars. 

Furthermore, we inspected the residual sum of squares (RSS) between the \swan's model and smoothed power spectrum of each star. This serves as a check of how well our model matches our data, so we can remove any poorly modeled stars. We removed any stars with RSS above 100 (ppm$^2/ \mu$Hz)$^2$; this filter removed a further 5 stars, or $\sim$0.04\% of the sample.
 
Figure \ref{ast_result} shows our final results for the seismic sample. We observe excellent agreement between the inferred and asteroseimic \logg, with a robust standard deviation\footnote{Robust standard deviation was calculated using \texttt{mad\_std()} from \texttt{Astropy} where \sigmamad $\approx$ 1.4826 MAD (median absolute deviation).} ($\sigma_{\textrm{mad}}$) of 0.02 dex with no significant bias for 14,003 stars, including outliers. Figure \ref{ast_result} demonstrates that \swan can successfully infer \logg from stellar power spectra to high precision across five orders of magnitude in surface gravity. Inferred \logg values and other stellar parameters for the seismic sample are provided in Table \ref{tb:astero_catalogue}.

Only 181 out of 14,003 stars, or $\sim$1.3 \% of the final sample shown in Figure \ref{ast_result} are classified as outliers, where we define outliers to be greater than 3$\sigma$ from the reference \logg label. To investigate these 181 outliers, we plotted normalized histograms to show stellar properties of outliers compared to stars with successful inference in the right panel of Figure \ref{ast_result}. Outliers peak at higher radii, lower effective temperatures, and lower surface gravities. This indicates that outliers in the seismic sample are evolved, high luminosity stars. %However, due to a lack of significant correlation between outliers and regions of parameter space there is no notable indication why inference for some stars is unsuccessful.

A possible reason for the presence of outliers in our inference could be that rotation modulation is mixed in with the granulation signal in the power spectrum. This would indicate that our removal of long-period variations from the light curve during initial data processing steps was not completely successful. For instance, applying a wide smoothing filter on $\sim$90-days of photometry on evolved stars has little effect on removing the long-periodic variations, while applying a narrow filter below 2 days removes the granulation signal that we want to measure. Additionally, a lack of label space coverage of the training objects within this particular parameter space could also contribute to unsuccessful inference for bright stars. For instance, only $\sim$5\% of the combined seismic and \gaia samples have high radii (917 seismic stars with radii above 30 \solradius) and $\sim$6\% have high luminosities (1104 seismic stars with luminosities above 200 \sollum). Therefore, for evolved stars or stars above 30 \solradius, \swan is not successful on one quarter of \kepler data, presumably due to our inability to measure granulation in the power spectrum, as well as a sparsely populated training set within this region.

\subsection{Gaia sample}
\label{sec:gaia_results}

\begin{figure}[t!]
\begin{center}
\includegraphics[width=0.5\textwidth,angle=0]{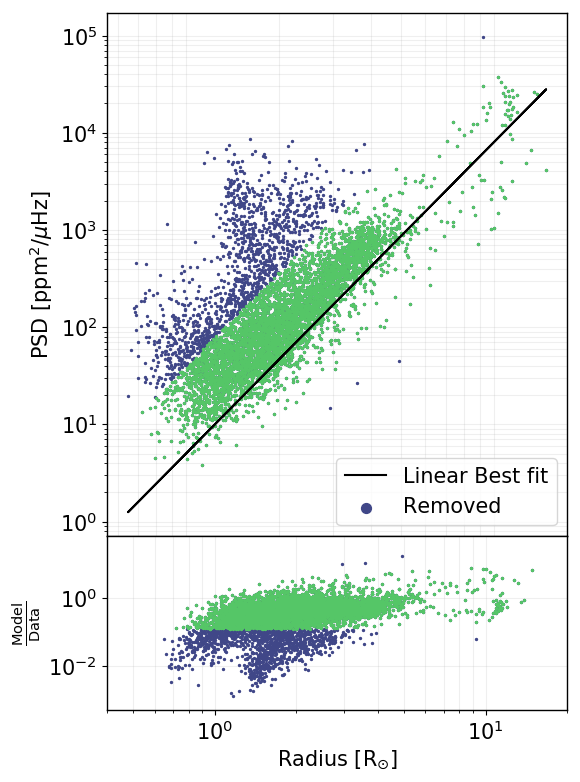}
\caption{Mean power spectral density between 10-12 $\mu$Hz as a function of stellar radius for stars in the \gaia sample. Stars dominated by variability other than granulation are identified as outliers which are shown as blue data points, while green data points represent stars within $\sim$7.9 ppm$^2/\mu$Hz of the best fit line.}
\label{pande_pr}
\end{center}
\end{figure}

\begin{figure*}[t!]
\begin{center}
\includegraphics[width=1\textwidth,angle=0]{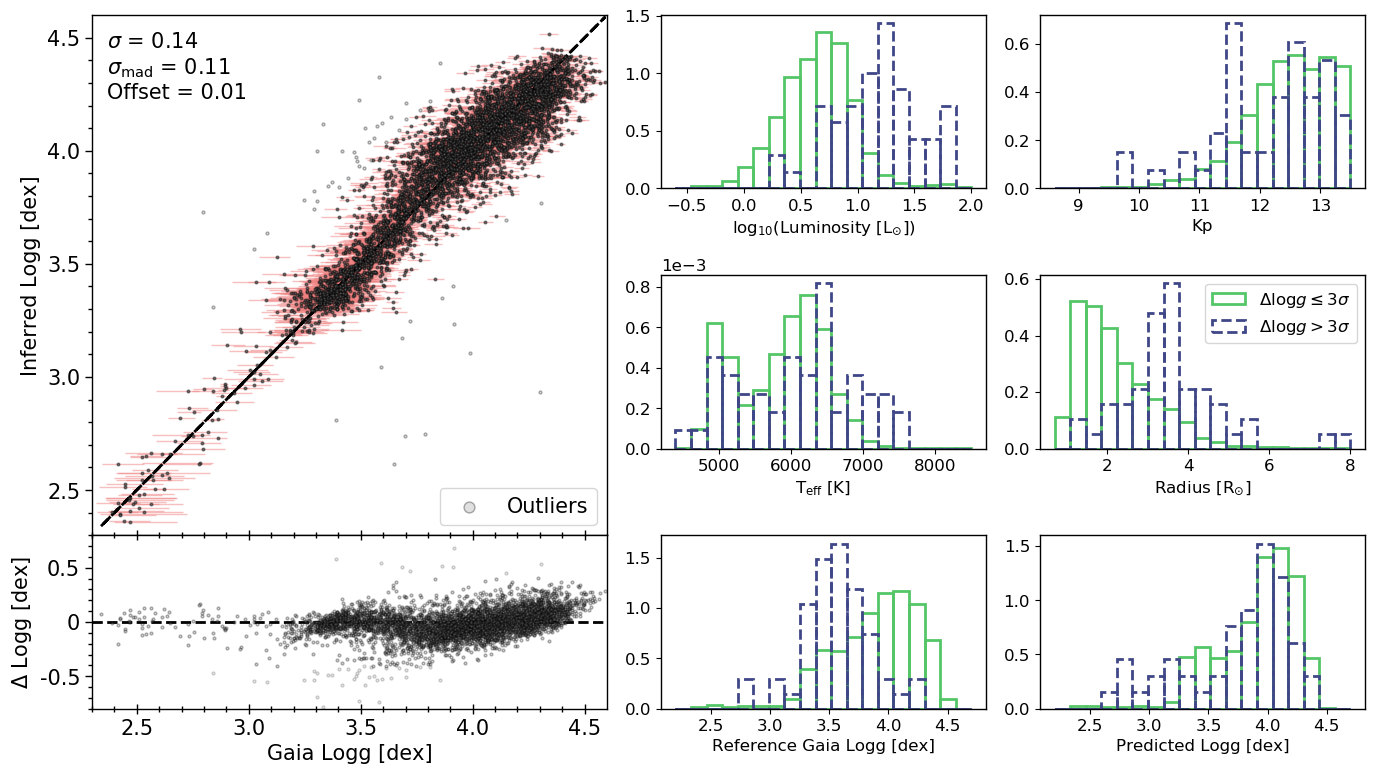}
\caption{\textit{Left}: Surface gravities for \gaia sample inferred with \swan, compared to the reference \logg values as derived in \cite{berger_2020}. The bottom left panel shows the difference between the reference \gaia \logg and the inferred \logg. The top left hand corner shows the precision and bias metrics of the inference, where $\sigma$ is the root mean square (RMS) and \sigmamad is the robust standard deviation for 4,697 stars (including outliers). Outliers, defined as stars with a difference between their reference and inferred \logg above $3\sigma=0.41$ dex, are shown as grey points, and errors in \gaia \logg are shown in red. \textit{Right}: Histograms demonstrating stellar properties of the \gaia sample. The green solid bars show normalized counts for non-outliers, and the purple dashed bars show properties of outliers.}
\label{pande_result}
\end{center}
\end{figure*}

We performed similar analysis for the seismic sample on the \gaia sample, which contains stars with \logg measured through isochrone fitting \citep{berger_2020}. Similar to the seismic sample, initial analysis of the \gaia sample showed outliers in the cross-correlation plot, corresponding to stars which have high power spectral density at low frequencies. Figure \ref{pande_pr} shows mean power spectral density between 10-12 $\mu$Hz as a function of stellar radius for the \gaia sample. Unlike Figure \ref{astero_pr}, Figure \ref{pande_pr} does not include the turnoff at $\sim$20 \solradius because the \gaia sample does not include many evolved stars. We modeled the distribution in Figure \ref{pande_pr} using Equation \ref{eq:pr_eq}, and removed stars with power density larger than $\sim$7.9 $\text{ppm}^2/\mu \text{Hz}$ from the best fit line. Through this filter, we removed $\sim$21\% of the sample, or 1,258 from an original 5,964 stars. We also removed stars with residual sum of squares above 100 (ppm$^2/ \mu$Hz)$^2$ which removed a further 9 stars.  

Figure \ref{pande_result} shows the comparison between the inferred \logg using local linear regression and reference \gaia \logg derived in \citet{berger_2020}, with residuals shown in the bottom left panel. The final cross-validation gives a robust standard deviation of 0.11 dex with no significant bias for 4,697 stars, including outliers. Inferred \logg values and other stellar parameters for this sample are provided in Table \ref{tb:pande_catalogue}. 

Figure \ref{pande_result} shows a fairly strong correlation but with a large scatter for stars with \logg above $\sim$3.5 dex. Only 51 out of 4,697 stars, or $\sim$1\% of the sample, are classified as outliers. Similar to the seismic sample, we explored the cause of outliers by plotting normalized histograms of stellar parameters representing outliers and the rest of the sample shown in the right panel of Figure \ref{pande_result}. Outliers peak at higher luminosities, higher radii and lower surface gravities. 

Some of the outliers could be due to pixel contamination, for example through an asteroseismic signal being blended into the \kepler aperture of the target stars \citep{hon_2019}. To explore this, we investigated contamination values of \kepler data from MAST, as well as the re-normalized unit-weight error (RUWE) value from \gaia \citep{lindegren_2018}. MAST provides a measure of contamination due to neighbouring stars where a value of 0 implies no contamination and a value of 1 implies light is all due to background. RUWE is the magnitude and colour-independent re-normalization of the astrometric $\chi^2$ fit in \gaia DR2, which is sensitive to close binaries \citep{gaia, evans_2018, lindegren_2018, berger_2020}. Out of the 61 outliers, only 15 stars have an RUWE above 1.2, which is a typical indicator of binarity. Out of the entire \gaia sample, 11\% of stars -- or 517 out of 4,697 -- have a RUWE above 1.2. Additionally, none of the outliers in our sample had a MAST contamination parameter above 0.027. We conclude that contaminating flux and stellar binarity from nearby stars are not driving our misestimated \logg values. 

We further investigated the differences between Figures \ref{ast_result} and \ref{pande_result}. The \logg values in the seismic sample are calculated using scaling relations which give very precise stellar parameters \citep{mathur_2017, yu_2018}. The \logg values in the \gaia sample have larger error bars and are calculated through isochrone fitting which chooses the most likely models based on input parameters and their uncertainties \citep{huber_2017, berger_2020}. For instance, \gaia parallax and 2MASS K magnitude provide a luminosity which are combined with a colour and isochrones to derive \logg. This produces less precise \logg values ($\sim$0.05 dex versus $<$0.03 dex from asteroseismology), which may additionally be affected by systematic errors from the use of a single isochrone grid. Therefore, we believe the success of \swan is highly dependent on the accuracy of the labels in the reference set, and that the difference in predictive uncertainty achieved for the seismic and \gaia samples is driven by the uncertainty of the training \logg values. 

\subsection{S/N Metric for Detecting Granulation}

\begin{figure}[t!]
\begin{center}
\includegraphics[width=0.5\textwidth,angle=0]{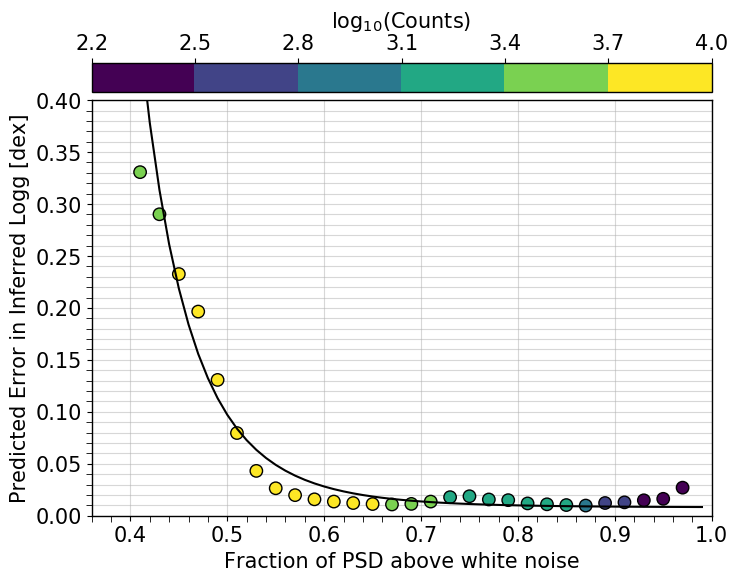}
\caption{Predicted uncertainty in inferred \logg as a function of fraction of power spectral density above white noise level. Absolute standard deviation was calculated using non-outliers in the seismic and \gaia samples for each 0.02 increment of signal-to-noise (S/N). The solid curve represents the best-fit to the data using a power law. The data points are coloured by the number of stars within the specified S/N range used to calculate predicted uncertainty.}
\label{inferred_logg_error}
\end{center}
\end{figure}

\begin{figure}[t!]
    \centering
    \includegraphics[width=0.5\textwidth]{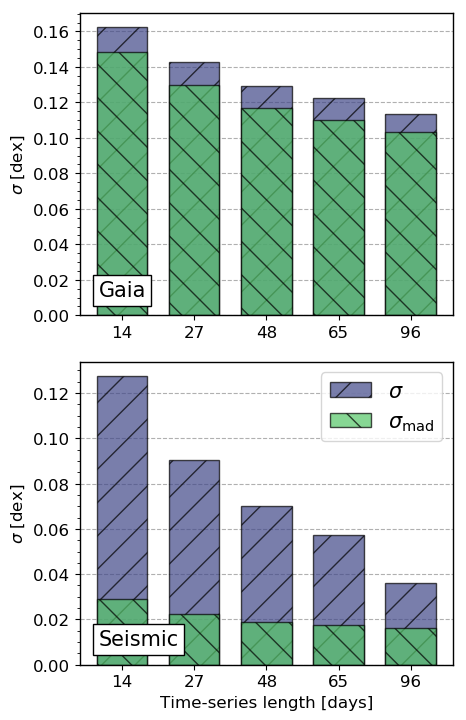}
    \caption{Recovered precision of surface gravity as a function of photometry length for the \gaia and seismic samples. The green and purple bars show the standard deviation and the robust standard deviation using median absolute deviation respectively.}
    \label{fig:lc_cut}
\end{figure}

To explore the source of the systematic differences in Figure \ref{pande_result}, we calculated a signal-to-noise (S/N) metric by counting the number of points in the spectrum with power above the white noise level as a fraction of the total number of points, resulting in a S/N range of $\sim$0.02 to 1.0 for the combined seismic and \gaia samples. 

To investigate the S/N value pertaining to a white noise dominated spectra, we generated over 12,000 `noisy' light curves and calculated the power above the white noise in the power spectrum, where the white noise level was chosen as the average power at frequencies between 270-277 $\mu$Hz. The median S/N from simulated light curves was 0.45 indicating that we can expect S/N of $\sim$0.45 or below for white noise dominated spectra where granulation cannot be measured. 

In fact, the comparison in Figure \ref{pande_result} improves when we apply a S/N cut to our inferred \logg values for the \gaia sample. As the value of S/N cut increases, we retain less of the sample, but the bias seen at higher \logg diminishes in Figure \ref{pande_result}. This improvement due to a cut in S/N supports our conclusion that stars with low S/N are too faint to detect granulation.

Indeed, our inferred \logg values are systematically lower than predicted by evolutionary models for cool main sequence stars such as K dwarfs. While small biases may be expected from systematic model errors (e.g. \citealp{boyajian_2013}), it is more likely that the lack of detectable granulation leads to an underestimation of \logg. This implies that applying granulation-based \logg to K dwarfs is only possible for the brightest stars, even in the \kepler sample.

\subsection{Uncertainties in Inferred \logg}
To derive uncertainties in our inferred \logg values, we applied \swan on both the seismic and \gaia samples by adding white noise to the photometry to investigate the scatter in inference as a function of S/N. To anticipate the noise embedded in the photometry, we simulated 10,000 sets of noisy data between 0.001-$10^4$ ppm and produced a relationship between time domain noise and white noise in the power spectrum. The resulting relation was combined with our previously derived relation between white noise level and \kepler magnitude (see Section \textbf{\ref{data_prep}}) to create a correlation between time domain noise and \kepler magnitude.

To quantify how the inference degrades for a noise dominated light curve, we applied \swan on photometry with added noise ranging from 0.5 to 6 times the time domain noise. The noise was only added to the test star and not the entire set of reference stars, and the remaining data were prepared following steps outlined in Section \textbf{\ref{data_prep}}.

We derived uncertainties in inferred \logg values by combining \swan's results from seismic and \gaia samples with many different amplitudes of noise added. Figure \ref{inferred_logg_error} shows the robust standard deviation (\sigmamad) in \logg inference as a function of S/N between 0.1-1.0, with S/N bin width of 0.02, coloured by the number of stars within the specified S/N range. For instance, a star with S/N between 0.50-0.52 would have a typical error of 0.08 dex on its inferred \logg measurement. We used the following equation to fit a relation between S/N and predicted error in inferred \logg values,
\begin{equation}
    \label{eq:inferred_err}
    \sigma = 3.39\cdot 10^{-4}s^{-8.06} + 6.75\cdot10^{-3} 
\end{equation}
where $s$ is the S/N and $\sigma$ is the predicted inferred \logg uncertainty in dex. Stars with S/N below 0.36 and above 0.96 were removed from the fitting routine due to small number of stars with S/N in this range. Equation \ref{eq:inferred_err} was used to assign errors to inferred \logg measurements in Tables $\ref{tb:astero_catalogue}$ and $\ref{tb:pande_catalogue}$. An uncertainty floor of 0.02 dex was set for any predicted uncertainties below 0.02 dex. 

In Figure \ref{inferred_logg_error}, the scatter increases for lower S/N values implying that a lack of granulation signal makes it difficult for \swan to make a successful inference. Therefore, we do not report a \logg measurement for any stars with S/N below 0.36 due to an unsuccessful inference. Stars with S/N above 0.96 were assigned an error of 0.2 dex since most stars with S/N in this range are high-luminosity giants for which one quarter of \kepler is not sufficient to successfully measure granulation. For future use of \swan, we recommend caution in derived \logg measurements for any star with S/N below 0.45 and radius above 30 \solradius.

\subsection{Photometry Baseline}

We explored the effect of photometry length on \swan's performance to estimate the precision achievable with data from other surveys. While \kepler delivered light curves at a baseline of roughly 90 days at 29.4 minute cadence, TESS provides time-series photometry at two cadences with a baseline ranging from 27 days to a full-year. Therefore, we tested our model on four photometry baselines of 14, 27, 48 and 65 days with no change in cadence. To process the light curve, we only selected data points if they occurred at days less than the specified baseline. Aside from truncating the light curves, we prepared the data as described in Section \ref{sec:data_prep}. 

Figure \ref{fig:lc_cut} shows the effect of shortened photometry length on the precision of recovered \logg using \swan on the seismic and \gaia samples. As expected, the scatter increases as the length of the light curve decreases, affecting both the root mean square (RMS) and the robust standard deviation (MAD). For the \gaia sample, the RMS scatter at 27 days is $\sigma=0.14$ dex, which is 23\% larger than the RMS scatter at 96 days (full length photometry). However, high precision, measured by \sigmamad = 0.11 dex, is still achieved despite the shorter baseline.

For the seismic sample, while the RMS scatter doubles at 27 days compared to full-length photometry (from $\sigma$ = 0.04 dex to 0.09 dex), the MAD precision is still less than 0.03 dex for the shorter baseline. Unlike the \gaia sample, there is a significant difference between the RMS and the robust standard deviation, where the RMS at 27 days (TESS-baseline) is over twice as large as the RMS at full-length photometry. Therefore, the use of \swan on shorter baseline data produces more outliers. These individual objects are not the same outliers when using the longer baseline ($\sim$90 days) data. As expected, higher quality data produces higher precision results which emphasizes the importance of a high quality training set to obtain high fidelity training labels. 

Nevertheless, we expect \swan to recover stellar parameters with an increased scatter for TESS-like data (27 days) compared to full \kepler light curves. However, the shorter baseline still produces precision comparable to, or even less than, data-driven methods found in literature \citep{casey_2016, ness_2015, ness_2016, ness_2018}. Therefore, \swan can be successfully applied to TESS photometry to infer stellar properties despite a shorter baseline. 

\section{A Comparison: inference of \logg using \cannon}
\label{sec:thecannon}

To investigate the inference of stellar surface gravity using \cannon \citep{ness_2018}, we use a single label second-order polynomial model that is built on our training set of data. Like \swan, this is data driven: a reference set of stars (training set) are used to infer labels for a new set of stars (test set). However, \cannon as implemented in \citet{ness_2018} is quite distinct from \swan, which builds a local model using nearest neighbours in the data space for every test object that is labeled. We briefly describe \cannon below and refer the reader to \cite{ness_2015} and \cite{ness_2018} for a more detailed discussion.

To use \cannon on time-series data, the data must satisfy the following conditions: (a) stars with the same labels must have the same spectra and (b) power density at a given frequency should vary smoothly with stellar labels.  To use \cannon, we converted our data to the frequency domain, by taking the Fourier Transform denoted by $g(f)$. We then smoothed over the power spectrum by 2 $\mu$Hz to remove sharp peaks produced by oscillation modes in the spectrum.

We take $n$ reference objects with their known labels $\ell$. The spectral model is characterized by a coefficient vector $\theta$ that allows the prediction of the amplitude, $G_{nf}$, at every frequency $f$ of the power spectrum for a given label vector:
\begin{equation}
    \begin{aligned}
    \ln p(G_{nf}|\theta^T_f,\ell_n,s^2_f)=&-\frac{1}{2}\frac{[G_{nf}-\theta^T_f \cdot \ell_n]^2}{s^2_f+\frac{\sigma^2_{nf}}{g_{nf}}}\\
    &-\frac{1}{2}\ln(s^2_f+\frac{\sigma^2_{nf}}{g_{nf}})
    \end{aligned}
\end{equation}

The noise term is a combination of uncertainty variance at each frequency and the intrinsic variance of the model of the fit, $s^2$. A term $1/g_{nf}$ is included to represent propagated amplitude error since the data are transformed to a logarithmic amplitude. At training time, the coefficient vector for the quadratic model is solved for at every frequency step. We solve for 3 coefficients at each frequency step, such that $\ell_n$=[1, \logg, \logg$^2$].

\begin{figure}[t!]
\begin{center}
\includegraphics[width=0.5\textwidth,angle=0]{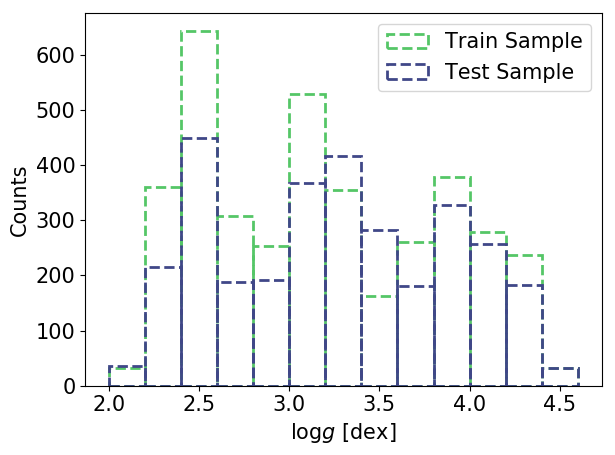}
\caption{Distribution of stellar surface gravity (\logg) used for training (3,692 stars) and testing (3,675) \cannon.}
\label{cannon_sample_hist}
\end{center}
\end{figure}

\begin{figure}[t!]
\begin{center}
\includegraphics[width=0.5\textwidth,angle=0]{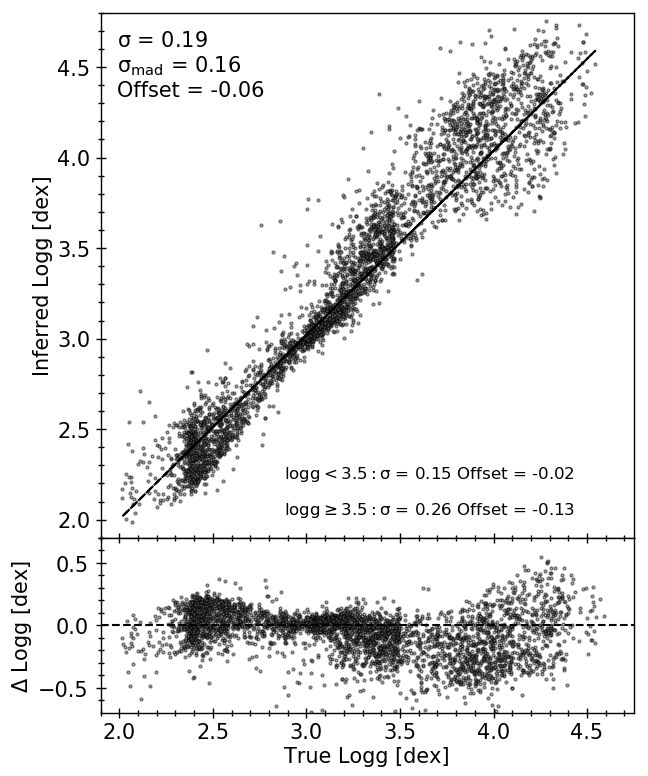}
\caption{Comparison of inferred surface gravities as a function of reference \logg using \cannon trained on \logg only. Reference \logg values were obtained from \citet{mathur_2017}, \citet{yu_2018} and \citet{berger_2020} for 3,675 stars.}
\label{cannon_results}
\end{center}
\end{figure}

\begin{figure*}[t!]
\begin{center}
\includegraphics[width=\textwidth,angle=0]{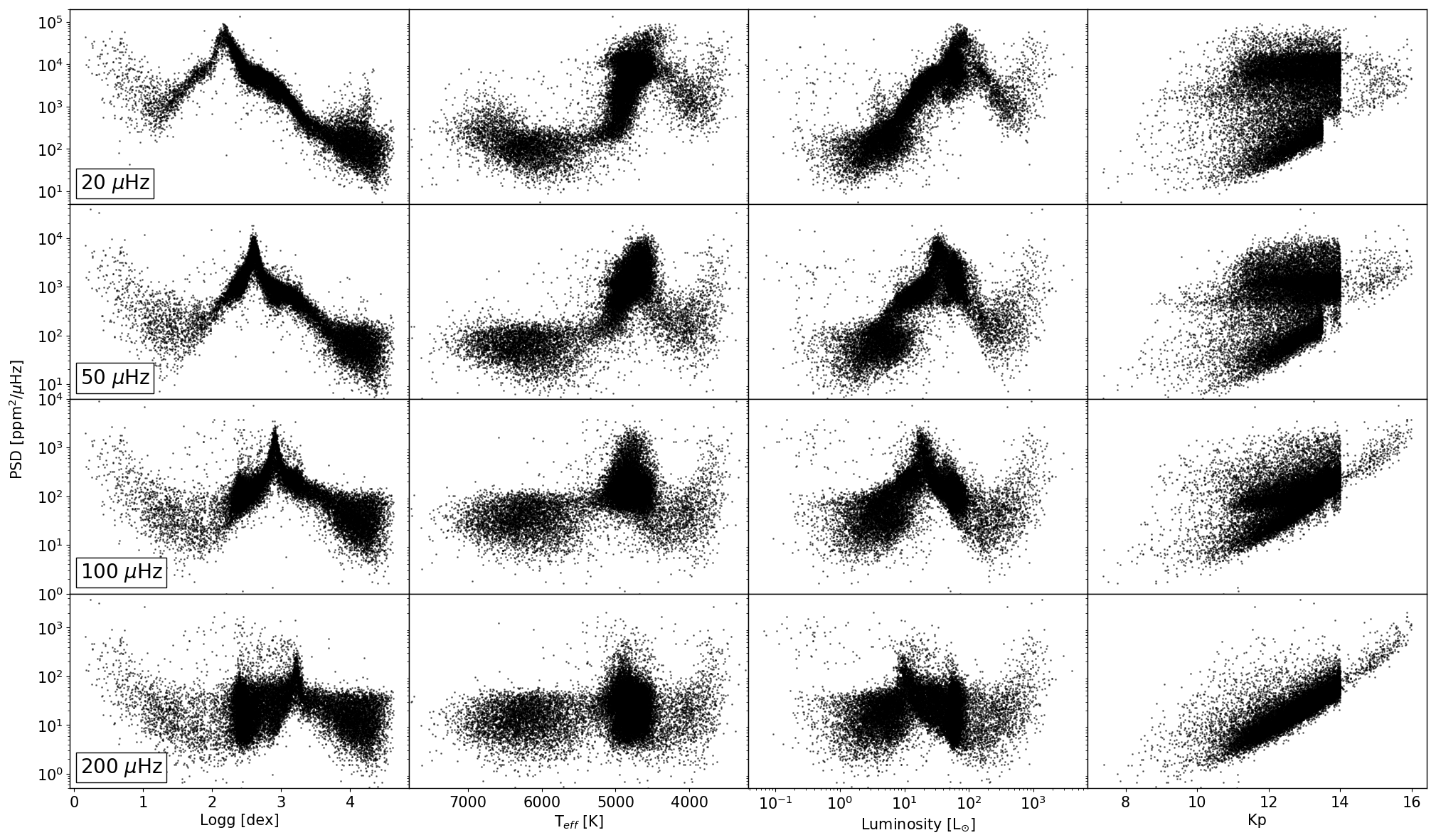}
\caption{Information content at individual frequencies in the power spectrum plotted for four different labels. Each row corresponds to the power spectral density at a single frequency (ie. 20, 50, 100 and 200 $\mu$Hz), and each column corresponds to a label that we attempted to infer (\logg, \teff, luminosity and Kp). This figure demonstrates that for the representative pixels shown, there are relationships between the data and label, but the relationship between the data and the label is not described by a polynomial model over the full range of the data. Over a narrow enough label range, the \logg shows a fairly tight linear relationship with the data amplitude, particularly at lower frequencies.}
\label{pixel_info}
\end{center}
\end{figure*}

Figure \ref{cannon_sample_hist} shows the \logg distribution of the 3,692 stars used in our training set and 3,675 stars used in our testing set. The training set is used to build the model and the test set is used to evaluate its performance in the label inference. The sample was split up in such a way to include sufficient number of stars in the training set to be representative of the parameter range, and contain at least a thousand dwarfs and giants in both the training and testing set. 

The training set used to create the model includes a total of 2,437 giants and 1,255 dwarfs (3,692 stars), and was tested on 2,450 giants and 1,225 dwarfs (3,675 stars) (Figure \ref{cannon_sample_hist}). All spectra used for training and testing the model were corrected for white noise following a linear relationship between white noise level and \kepler magnitude. 

Figure \ref{cannon_results} shows the inference of stellar surface gravity using \cannon on 3,675 stars with a \logg range of 2.0-4.6 dex. The top panel shows \cannon's performance compared to the reference label value and the bottom panel shows the residuals. When trained on \logg, \cannon infers \logg to a precision of 0.16 dex, but with a bias of -0.06 dex. While \cannon is successfully inferring \logg from the power spectra of light curves, the scatter and bias vary significantly as a function of \logg. For instance, it performs worse for subgiants and dwarfs with a scatter of 0.26 dex and a bias of -0.13 dex, and over-predicts \logg for 72\% of stars with \logg above 3.5 dex. 

Our results using \cannon with a second order polynomial model demonstrate that while effective across the red giant branch, this approach can not capture well the relationship between time domain variability and the \logg label over a large range of \logg values. A choice of a more flexible model for \cannon, like a Gaussian process, would possibly circumvent this problem, but at some  computational expense, and have lesser interpretability than a (global or local) linear or polynomial modeling approach. 

To examine the impact of our choices in \cannon's implementation, we trained the model using label combinations such as effective temperature ($T_\text{eff}$), luminosity ($L$), Kp and radius. While the model trained on luminosity was slightly successful, the inference failed for the rest of the labels. We also experimented with hyper-parameters such as the number of stars in the reference set, the region and length of the power spectrum used for inference, the region and length of the raw light curve, and specific regions of parameter space. \cannon is best at predicting \logg when the entire stellar power spectra is used, but we found that it could not match the performance of \swan.

To understand the better performance of \swan compared to \cannon, we examined the correlation between the power spectrum's amplitude of our reference objects and our labels, including \logg, \teff, $L$ and Kp. Figure \ref{pixel_info} shows that there are correlations between these labels and the power spectrum amplitude at different frequencies, and that each frequency shows a different label-power amplitude relation. It is clear that these correlations can not be captured with a polynomial model, particularly for the labels of \teff, $L$ and Kp. A local linear model on the other hand, implemented in \swan, can effectively capture the locally changing relationship between \logg and PSD, over the full range of the data. While the label-PSD correlation is tighter at lower frequencies for \logg in Figure \ref{pixel_info} (and note for comparison, is tighter at higher frequencies for Kp), there is information distributed across the full frequency range shown, including up to 200 $\mu$Hz, that can be used to infer \logg. We conclude that while \cannon is a useful method to derive granulation-based \logg for giants, a polynomial model is not sufficiently flexible to model the relationship between the data, and \logg or other labels that describe it, which causes most significant failures in the dwarf and subgiant regime.

\begin{figure}[t!]
\begin{center}
\includegraphics[width=0.5\textwidth,angle=0]{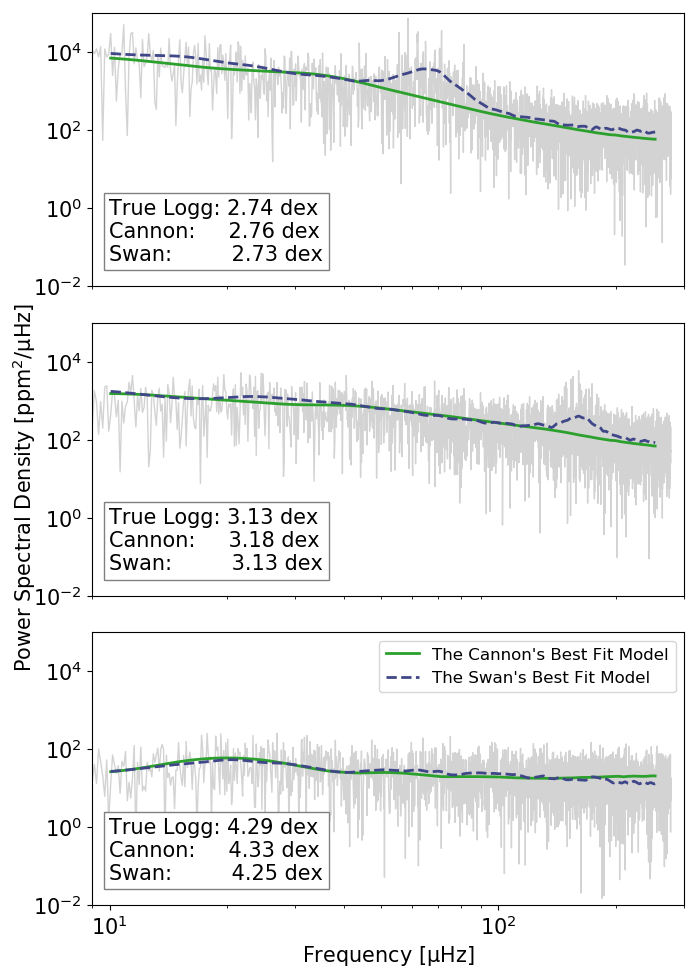}
\caption{Comparison of \cannon's best fit model in green and \swan's best fit model in dashed blue for three stars with varying \logg. The white noise corrected power spectrum of each star is plotted in grey.}
\label{cannon_vs_LLR}
\end{center}
\end{figure}

Figure \ref{cannon_vs_LLR} shows additional examples of comparing the \swan to \cannon for three stars with different \logg values. For each test star, \swan's model captures more of the variance in the data than \cannon's global model; it is able to reproduce the variations in the spectral shape in more detail, thus successfully modelling the oscillations when visible. In the bottom panel of Figure \ref{cannon_vs_LLR}, we observe that both \swan and \cannon's inferences are close to the reference \logg value for the dwarf star, but \cannon recovers a systematically higher \logg; the polynomial model does not label the data as effectively as \swan over the full parameter range of the data. 

Figure \ref{cannon_vs_LLR} shows that \swan works better at modeling the power spectra to derive granulation-based \logg measurements. In general, if \cannon's model is sufficient to describe the data, it should result in very similar results to a local linear modeling approach. Indeed, the main benefit of \cannon as used in \citet{ness_2018} compared to a local linear modeling approach may simply be in the lower computational expense, as nearest neighbours do not need to be evaluated and only one model rather than a unique model for each inference, is required.

\section{Comparison to Literature}
\begin{deluxetable*}{ccccc}
\tabletypesize{\footnotesize}
    \tablecolumns{5}
    \tablewidth{0pt}
    \tablecaption{\label{tb:lit} Summary of recovered \logg precision for different methods used to infer stellar surface gravity from photometric and spectroscopic data.}
    \tablehead{ \colhead{\logg Range [dex]} & \colhead{Precision [dex]} & \colhead{Reference} & \colhead{Data Type} & \colhead{Technique}}
    \startdata
    1.0-4.0 & 0.11-0.22 & \cite{ness_2015} & Spectroscopy & \cannon\\
    1.0-3.5 & 0.07 & \cite{ness_2016} & Spectroscopy & \cannon\\
    1.0-3.9 & 0.07 & \cite{casey_2016} & Spectroscopy & \cannon\\
    0.5-4.0 & 0.10 & \cite{ho_2017a} & Spectroscopy & \cannon\\
    1.0-3.0 & 0.14 & \cite{ting_2017} & Spectroscopy & \textit{The Payne}\\
    0.8-4.0 & 0.06 & \cite{fabbro_2017} & Spectroscopy & convolution neural networks \\
    0.0-5.0 & 0.07 & \cite{xiang_2019} & Spectroscopy & \textit{Data-Driven Payne}\\
    0.0-5.0 & 0.05 & \cite{ting_2019} & Spectroscopy & \textit{The Payne}\\
    1.2-4.8 & 0.12-0.2 & \cite{wheeler_2020a} & Spectroscopy & \cannon\\
    1.0-3.6 & 0.07 & \cite{tawny_2020} & Spectroscopy & \cannon \\
    0.5-5.0 & 0.06 & \cite{guiglion_2020} & Spectroscopy & convolution neural networks \\
    2.7-4.8 & 0.10 & \cite{rice_2020} & Spectroscopy & \cannon \\
    0.5-5.5 & 0.09 & \cite{zhang_2020} & Spectroscopy & support vector regression \\\hline
    2.5-4.6 & 0.10-0.20 & \citet{bastien_2013, bastien_2016} & Photometry (LC) & Flicker \\
    1.0-4.5 & 0.04-0.18 & \citet{kallinger_2016} & Photometry (ACF) & time-scale \\
    0.1-4.6 & 0.04-0.10 & \cite{bugnet_2018} & Photometry (PS) & random forest \\
    2.0-4.8 & 0.05 & \cite{pande_2018} & Photometry (PS) & time-scale \\
    1.5-3.3 & 0.10 & \cite{ness_2018} & Photometry (ACF) & \cannon\\
    0.2-4.6 & 0.02-0.11 & THIS WORK & Photometry (PS) & local linear regression \\\hline
    \enddata
    \tablecomments{LC: light curve; PS: power spectrum; ACF: auto-correlation function}
    
\end{deluxetable*}

\subsection{Spectroscopy}
Data-driven methods have been employed on large surveys of spectroscopic data to infer stellar labels, including \teff, \logg, mass, [Fe/H], and detailed chemical abundances \citep[e.g.][]{ness_2016, buder_2018, leung_2019, behmard_2019, birky_2020, galgano_2020, tawny_2020, rice_2020}. Simple data-driven methods, like \cannon,  are particularly useful for `label-transfer' where a high-resolution survey can be used to label a low-resolution survey to high precision compared to prior approaches \citep[e.g.][]{casey_2016, ho_2017b, wheeler_2020a}. The high-resolution APOGEE survey (R=22,500) delivers \logg to a precision of $\sim$0.02 dex \citep{apogee_dr16} with $S/N > 100$, using comparisons to stellar models with their \textsc{aspcap} pipeline \citep{aspcap_2015}. Using the data-driven approach of \cannon, a high fidelity \textsc{aspcap} sample has been used to label lower-resolution RAVE (R=8000) stars and LAMOST (R=1000) stars to a precision of $\sim$0.04 dex and $\sim$0.10 dex respectively using stars in common between the surveys \citep{casey_2016, ho_2017a}. 

Further data-driven approaches to infer stellar labels includes convolutional neural networks trained on RAVE DR6 and APOGEE DR13 spectra to recover \logg to a precision of 0.06 dex \citep{fabbro_2017, guiglion_2020}, and Bayesian artificial neural networks with dropout variation inference trained on APOGEE DR14 data to recover \logg and elemental abundances to $\sim$0.02 dex despite a low SNR (SNR$\sim$50) \citep{leung_2019}. With a non-parametric, neural-net-like functional form, \textit{The Payne} fits physical spectral models to recover stellar parameters and chemical abundances to high precision \citep{ting_2017, ting_2019}, and can be combined with \cannon's data-driven approach to derive stellar labels (\textit{Data-Driven Payne}, \citealp{xiang_2019}). Other methods to recover stellar labels from spectroscopic data include line-by-line differential analysis on high resolution (R=115,000) spectra \citep{bedell_2018}, and support vector regression (SVR) trained on LAMOST and APOGEE spectra \citep{ zhang_2020}.

Compared to spectroscopy, the photometric method presented here provides better precision over a larger range of evolutionary states (see Table \ref{tb:lit}). This is consistent with the fact that \logg information in spectral lines is typically degenerate with other atmospheric parameters such as \teff and [Fe/H] \citep{torres_2012}, while amplitudes of photometric granulation are predominantly determined by surface gravity \citep{kjeldsen_2011}.

\begin{figure}[t!]
\begin{center}
\includegraphics[width=0.5\textwidth,angle=0]{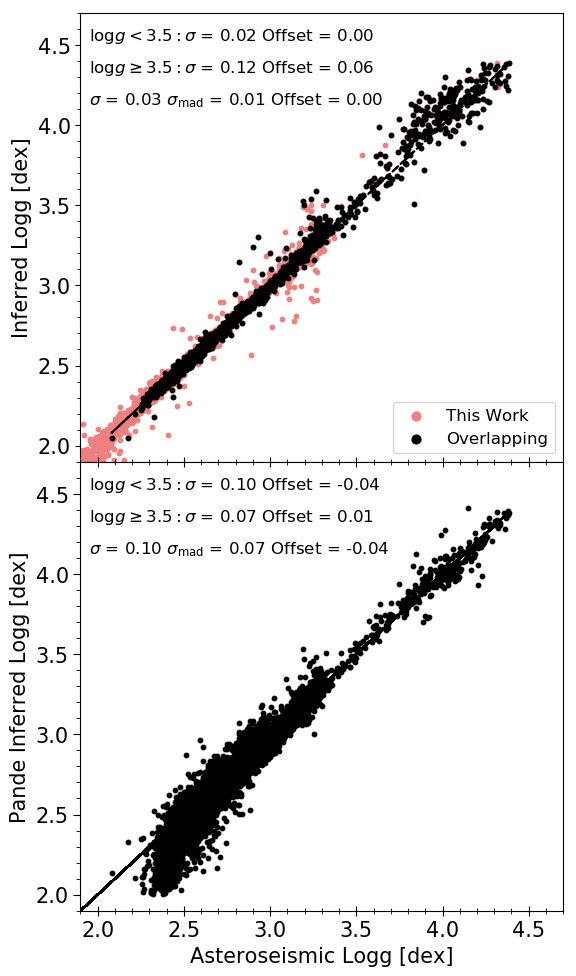}
\caption{Comparison of \swan's performance to the results of \cite{pande_2018} to predict \logg from power spectra for 5,429 stars. \textit{Top}: Inferred \logg from \swan versus asteroseismic \logg. The red points show the complete seismic sample, and the black points show overlapping stars found in both our seismic sample and \cite{pande_2018}. \textit{Bottom}: Cross-validation of results from \cite{pande_2018} versus asteroseismic \logg.}
\label{pande_vs_us}
\end{center}
\end{figure}

\begin{figure*}[t!]
\begin{center}
\includegraphics[width=\textwidth,angle=0]{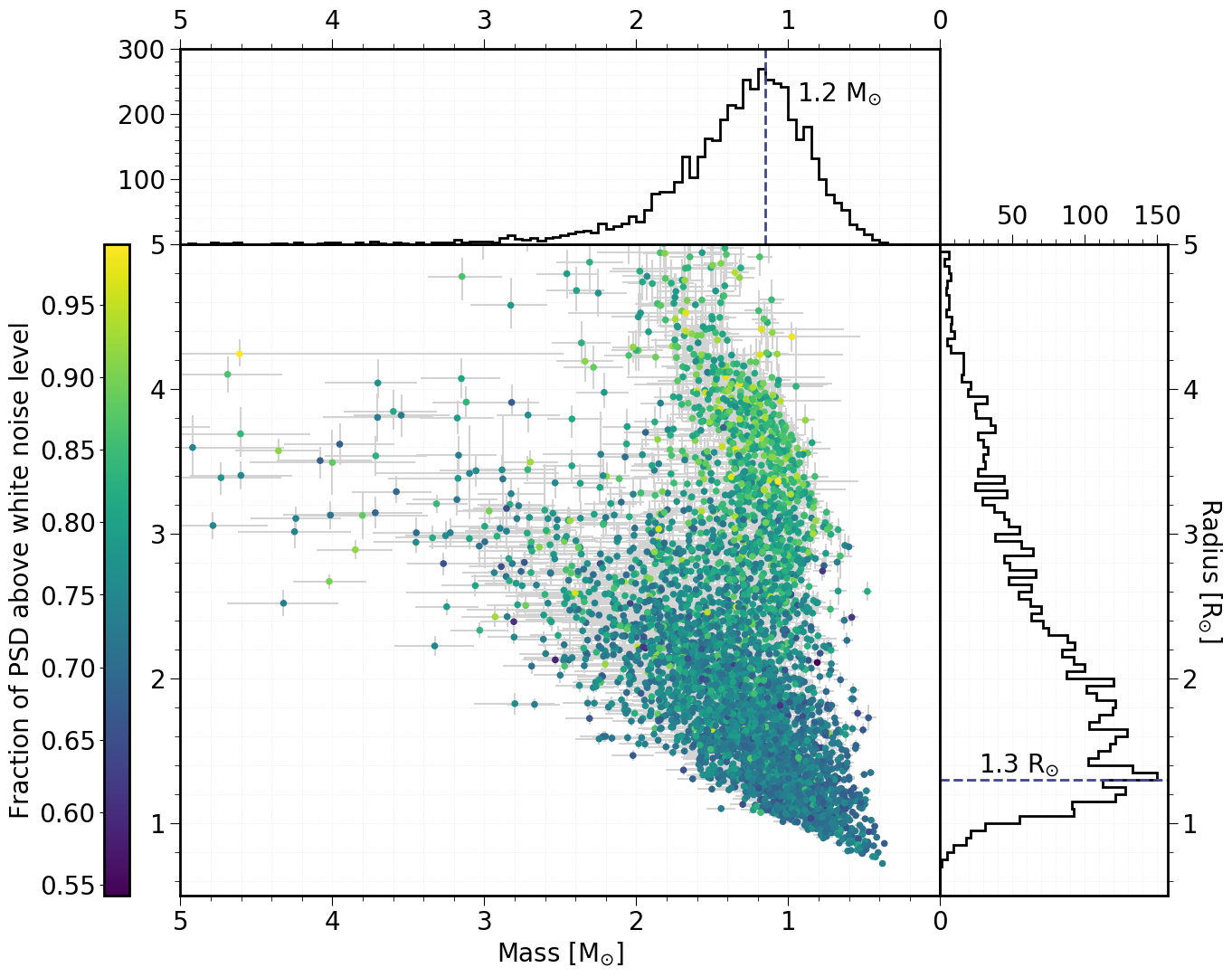}
\caption{The distribution of masses and radii for \gaia sample using inferred \logg from this work and \gaia derived radii from \cite{berger_2020} to calculate mass. Mass uncertainties are calculated using standard deviation of inferred \logg values and radii errors from \cite{berger_2020}. The stars are colour-coded by the fraction of power spectral density above white noise level. There are an additional 111 stars outside of the axis limits.}
\label{mass_radius}
\end{center}
\end{figure*}

\subsection{Photometry}

Using light curves of \kepler stars with stellar surface gravity between 2.5-4.6 dex, \citet{bastien_2013, bastien_2016} inferred \logg from power spectra to $\sim$0.1-0.2 dex precision, by employing the ``Flicker'' ($F_8$) method, which measures stellar variations on time-scales of 8 hours or less, and using asteroseismic samples for calibration \citep{bruntt_2012, thygesen_2012, stello_2013, huber_2013, chaplin_2014}, but restricted stars with \logg $>$ 2.7 dex. Although they recovered \logg for 27,628 stars, the $F_8$ method suffers from a degeneracy where highly evolved stars, or stars with lower \logg, may appear as subgiants. Additionally, they find that systematic uncertainties of overlapping red giants-clump stars increase to 0.2 dex; therefore, a red giant with $F_8$-inferred \logg between 2.9-3.2 dex may in fact have a true \logg that is $\sim$0.3 dex lower \citep{bastien_2016}.

\citet{kallinger_2016} used typical brightness variations time-scales in \kepler data to infer stellar surface gravities to 4\%, by utilizing a self-adapting filter which isolates the gravity-dependent signal from instrumental effects. They employed the auto-correlation function to determine the typical time-scale of the residual signal, and used asteroseismically derived surface gravities across a wide-range of stellar masses and evolutionary states to calibrate their relation. Although they achieved \logg precision of $\sim$0.04-0.18 dex for 1,200 stars, a limitation of their technique is that the time-scale method requires sufficient temporal resolution. Therefore, recovering \logg using their technique on main sequence stars requires short cadence data which are limited given current surveys.

Data-driven approaches have been previously used to determine properties from time-series photometry \citep[e.g.][for red giant stars]{ness_2018}. Our use of local linear modeling is a marginally more computationally expensive approach compared to \cannon as used in \citet{ness_2018}, but the model flexibility in describing the relationship between time domain data and their labels enables determining stellar properties across evolutionary states.

A recent method to derive \logg from time-series observations was demonstrated in \citet{pande_2018} who derived granulation-based \logg values using the power spectrum rather than the light curve or the auto-correlation function. They measured the granulation background in a limited region of the power spectrum centered at 0.08\numax with a fractional width of 20\%. For all quarters of \kepler data, they calculated the median granulation power in this region, converted to \numax and calculated \logg given a linear relationship between \logg and \numax \citep{kjeldsen_bedding}. 

In the top panel of Figure \ref{pande_vs_us}, we compare the results of our approach on the high fidelity asteroseismic stars within the \citet{pande_2018} sample. The bottom panel shows the performance of the \citet{pande_2018} pipeline itself against these reference asteroseismic objects. For 5,429 stars found in both samples, \swan performs better than the methodology by \citet{pande_2018}, with a robust standard deviation of 0.03 dex compared to \sigmamad = 0.10 dex using their method. For subgiants and dwarfs, or stars with \logg $\geq$ 3.5 dex, their analysis yields a marginally better RMS scatter of 0.07 dex compared to our RMS scatter of 0.12 dex. For giants, \swan performs better with a RMS scatter of 0.02 dex compared to their method which yields a RMS scatter of 0.10 dex. 

Interestingly, the bottom panel of Figure \ref{pande_vs_us} shows that \cite{pande_2018} predict a wide range of \logg for a narrow range of giants, seen as a turnoff at \logg $\approx$ 2.4 dex. They used only a part of the power spectrum for their inference, centered at 0.08\numax with a fractional width of 20\%. Maximum oscillation power for giants is found at lower frequencies; therefore, limiting the spectrum might have made it difficult to differentiate between giants with a range of \logg values. An advantage of our model compared to \citet{pande_2018} is that we use the full spectrum for inference, which allows us to capture granulation over a wide range of evolutionary states. Figure \ref{pande_vs_us} demonstrates that using a restricted region of power spectra prevents successful inference of surface gravity especially for high-luminosity giants. Conversely, we suspect that using the full power spectrum may introduce additional scatter for less evolved stars, for which a smaller fraction of the power spectrum is dominated by granulation.

\section{Stellar Masses}

Granulation-derived surface gravities in combination with radii from \gaia allow the calculation of stellar masses with little model dependence \citep{stassun_2017}. Therefore, predicting \logg using \swan enables us to calculate masses for thousands of stars by combining our inferred \logg with \gaia radius \citep{berger_2020},
\begin{equation}
    \text{g}=\text{GM}/\text{R}^2
\end{equation}
where g is the surface gravity in CGS, R is the stellar radius and G is the gravitational constant. 

In Figure \ref{mass_radius}, we show the distribution of derived masses for the \gaia sample, where the mass ranges from $\sim$0.4-5.3 \solmass with a mass peak at 1.2 \solmass and a radius peak at 1.3 \solradius. To calculate mass uncertainties, we used Figure \ref{inferred_logg_error} to assign an error to our inferred \logg values while radius uncertainties were obtained from \citet{berger_2020}. 
The mass uncertainty ranges from $\sim$5-47\% with a median of $\sim$7\%. Only $\sim$1\% of stars have a fractional mass uncertainty above 20\%. There are 16 stars above 4 \solmass which all have a difference between their reference and inferred \logg above 0.33 dex. In fact, $\sim$83\% of stars that have a derived mass above 3.5 \solmass have a difference between their reference and inferred \logg above 0.25 dex. Such high mass stars are not expected to have convective envelopes, and thus our mass values are very likely overestimated. 

In Figure \ref{mass_radius}, we find a low number of small, high mass stars (bottom left region) since these stars would be rare and found in the post main sequence phase. The curved branch at $\sim$1 \solmass for radii above $\sim$2.4 \solradius represents subgiant stars. We find a higher density of stars around solar mass and solar radius, which is expected since \kepler was designed to observe solar-type stars \citep{kepler_borucki, kepler_koch}. Inferred masses including uncertainties for both the seismic and \gaia samples are provided in Tables \ref{tb:astero_catalogue} and \ref{tb:pande_catalogue}, respectively.

\section{Conclusions}
\label{sec:summary}
We presented a simple but effective data-driven approach to predict stellar surface gravity from power spectra of time-series photometry, which we term \swan. Our main results are as follows:
\begin{itemize}
    \setlength{\itemsep}{0pt}%
    \item We demonstrated that local linear modeling of time-series power spectra can recover stellar surface gravities (\logg) to a precision of \sigmamad = 0.02 dex for \logg between 0.2-4.4 dex, when implemented using a reference sample of stars with astroseismically derived \logg (top panel of Figure 1). \swan infers \logg from the full frequency range of power spectra, and delivers comparable or better precision than previous approaches on time-domain data \citep[e.g.][]{pande_2018}, and spectroscopic data, using e.g. \cannon \citep{ness_2015, ness_2016, casey_2016, ho_2017b, ho_2017a}. \swan maintains high performance across a wide range of evolutionary states. 
    
    \item We used \swan to infer stellar surface gravities to a precision of \sigmamad = 0.11 dex, using a reference set of stars with \logg labels derived from isochrone fitting in \citet{berger_2020} to gain larger coverage across evolutionary states (see bottom panel of Figure 1). Using these \gaia derived \logg values leads to a higher scatter and systematic biases for dwarfs compared to reference objects with high precision asteroseismic \logg values. The cause for the higher scatter in this sample could be due to insufficient information for faint, higher \logg stars. Inspection of the recovered fraction as a function of a signal-to-noise metric, or the fraction of power spectral density above white noise level, demonstrated that a lack of detectable granulation leads to an inaccurate \logg measurement. Therefore, granulation-based \logg is likely applicable to very few K dwarfs, and we recommend caution for derived \logg values for any star with low signal-to-noise (faint dwarfs), or stellar radius above 30 \solradius (high-luminosity giants). 

    \item We derived empirical masses for 4,646 subgiant and main sequence stars by combining our inferred \logg values with \gaia radii and uncertainties from \citet{berger_2020}. We find an average mass of 1.2 \solmass representative of the \kepler sample, with a typical uncertainty of 7\%. The sample increases the number of model-independent masses over the asteroseismic sample by a factor of 10. 
    
    \item We tested the effect of photometry baseline on the precision of recovered \logg using \swan. We demonstrate that the precision remains high ($\sigma <$ 0.15 dex for both samples and \sigmamad $<$ 0.03 dex for the seismic sample) even for 27 day long time-series, which implies that the method can be applied to stars observed by TESS which have sufficiently high enough S/N.

    \item We also applied \cannon to power spectra of 3,675 stars and inferred \logg to a precision of \sigmamad = 0.16 dex, a poorer performance compared to \swan. We find that \cannon's use of a polynomial model is not sufficiently flexible to model the relationship between the input data (power spectrum) and \logg over a wide range of evolutionary states.
\end{itemize}

\swan code is publicly available at \url{https://github.com/MaryumSayeed/TheSwan} and is archived on Zenodo\footnote{In this work, we use v1.0.0 of \swan code permanently archived on Zenodo at \url{https://doi.org/10.5281/zenodo.4422101}.}. The methodology presented here can be readily applied to other time-domain surveys such as TESS \citep{tess} which observes a much larger number of stars than \kepler. A challenge will be the existence of a sufficiently large reference set, which could be overcome through creation of synthetic light curves, or the detection of oscillations in a sufficient number of dwarfs and subgiants with TESS \citep{schofield_2019}.  We also plan to update the model to expand the number of labels we can use for inference from granulation such as stellar metallicity \citep{corsaro_2017, tayar_2019}, explore which combinations of labels result in a successful inference for other fundamental stellar parameters, and investigate the effect of choosing neighbours on the basis of other stellar parameters. Additional future work includes applying \swan on light curves from ground-based telescopes like ASASS-N \citep{asassn, auge_2020} and developing similar analyses for stars that exhibit other types of variability (ie. Delta Scutis, RR Lyrae and Gamma Doradus pulsators). 

\section*{Acknowledgments}
We are thankful to our anonymous referee for helpful comments which improved this manuscript. We gratefully acknowledge the tireless efforts of everyone involved with the \kepler and \gaia missions. We are grateful to Travis Berger for sharing his \textit{Gaia}-\kepler stellar properties catalog prior to publication and for useful discussions. M.S. would like to thank Kirsten Blancato, Jamie Tayar, Connor Auge, Vanshree Bhalotia, Casey Brinkman, Erica Bufanda, Ashley Chontos, Zachary Claytor, Sam Grunblatt, Nicholas Saunders, Lauren Weiss and Jingwen Zhang for helpful feedback and discussions. A.W. would like to thank Bridget Ratcliffe for helpful feedback on the manuscript.

The research was supported by the Research Corporation for Science Advancement through Scialog award No. 26080. D.H. acknowledges support from the Alfred P. Sloan Foundation and the National Aeronautics and Space Administration (80NSSC19K0597), and the National Science Foundation (AST-1717000). A.W is supported by the National Science Foundation Graduate Research Fellowship under Grant No. 1644869. M.N. is in part supported by a Sloan Foundation Fellowship. 

This research was partially conducted during the Exostar19 program at the Kavli Institute for Theoretical Physics at UC Santa Barbara, which was supported in part by the National Science Foundation under Grant No. NSF PHY-1748958.

This paper includes data collected by the \kepler mission and obtained from the MAST data archive at the Space Telescope Science Institute (STScI). Funding for the \kepler mission is provided by the NASA Science Mission Directorate. STScI is operated by the Association of Universities for Research in Astronomy, Inc., under NASA contract NAS 5–26555.

\software{\texttt{astropy} \, \citep{astropy}, \, \texttt{Matplotlib} \, \citep{matplotlib}, \, \texttt{NumPy} \, \citep{numpy}, \, \texttt{Pandas} \, \citep{pandas}, \, \texttt{SciPy} \, \citep{scipy}}, \texttt{The Cannon} \citep{ness_2015, ness_2018}

\begin{deluxetable*}{cccccccccc}[t!]
%\tabletypesize{\footnotesize}
    \tablecolumns{10}
    \tablewidth{0pt}
    \tablecaption{\label{tb:astero_catalogue} Stellar properties and output parameters for stars in the asteroseismic sample including derived $\log g$ and mass.}
    \tablehead{\colhead{KIC ID} & \colhead{K$_{\text{p}}$} & \colhead{T$_{\text{eff}}$ [K]} &  \colhead{Radius [$\text{R}_\odot$]} & \colhead{$\log g$ [dex]} & \colhead{$\log g$ Inferred [dex]} & \colhead{True Mass [$\text{M}_\odot$]} & \colhead{Inferred Mass [$\text{M}_\odot$]} & \colhead{SNR} & \colhead{Outlier Flag} }
    \startdata
    892738 & 11.73 & 4534 & $22.63^{+1.38}_{-1.19}$ & $1.77^{+0.02}_{-0.02}$ & $1.79^{+0.02}_{-0.02}$ & $1.15^{+0.21}_{-0.21}$ & $1.18^{+0.15}_{-0.13}$ & 0.82 & 0\\
    892760 & 13.23 & 5188 & $9.63^{+0.64}_{-0.52}$ & $2.39^{+0.01}_{-0.01}$ & $2.40^{+0.02}_{-0.02}$ & $1.00^{+0.14}_{-0.14}$ & $0.93^{+0.13}_{-0.11}$ & 0.82 & 0\\
    893214 & 12.58 & 4728 & $11.30^{+0.66}_{-0.63}$ & $2.52^{+0.01}_{-0.01}$ & $2.52^{+0.02}_{-0.02}$ & $1.54^{+0.09}_{-0.09}$ & $1.53^{+0.19}_{-0.18}$ & 0.81 & 0\\
    893233 & 11.44 & 4207 & $28.47^{+2.00}_{-1.86}$ & $1.67^{+0.01}_{-0.01}$ & $1.68^{+0.02}_{-0.02}$ & $0.86^{+0.09}_{-0.09}$ & $1.40^{+0.21}_{-0.19}$ & 0.76 & 0\\
    1026084 & 12.14 & 5072 & $12.79^{+0.70}_{-0.72}$ & $2.53^{+0.01}_{-0.01}$ & $2.58^{+0.02}_{-0.02}$ & $1.75^{+0.18}_{-0.18}$ & $2.27^{+0.27}_{-0.28}$ & 0.82 & 0\\
    1026309 & 10.60 & 4514 & $21.86^{+0.88}_{-0.84}$ & $2.12^{+0.02}_{-0.02}$ & $2.08^{+0.02}_{-0.02}$ & $2.71^{+0.42}_{-0.42}$ & $2.13^{+0.20}_{-0.19}$ & 0.80 & 0\\
    1026326 & 13.26 & 5123 & $7.11^{+0.26}_{-0.26}$ & $2.90^{+0.01}_{-0.01}$ & $2.90^{+0.02}_{-0.02}$ & $1.30^{+0.08}_{-0.08}$ & $1.54^{+0.13}_{-0.13}$ & 0.86 & 0\\
    1026452 & 12.94 & 5089 & $11.99^{+0.81}_{-0.47}$ & $2.46^{+0.01}_{-0.01}$ & $2.46^{+0.02}_{-0.02}$ & $1.58^{+0.17}_{-0.17}$ & $1.40^{+0.20}_{-0.13}$ & 0.79 & 0\\
    1027110 & 12.10 & 4190 & $26.06^{+1.81}_{-1.73}$ & $1.70^{+0.01}_{-0.01}$ & $1.68^{+0.02}_{-0.02}$ & $1.15^{+0.13}_{-0.13}$ & $1.15^{+0.17}_{-0.16}$ & 0.79 & 0\\
    1027337 & 12.11 & 4671 & $8.44^{+0.32}_{-0.32}$ & $2.77^{+0.01}_{-0.01}$ & $2.78^{+0.02}_{-0.02}$ & $1.34^{+0.07}_{-0.07}$ & $1.58^{+0.14}_{-0.14}$ & 0.88 & 0\\
    \enddata
    \tablecomments{Stellar effective temperatures and radii (including uncertainties) were obtained from \cite{berger_2020}. Surface gravity, true stellar mass and respective uncertainties were obtained from \cite{mathur_2017} and \cite{yu_2018}. Inferred mass uncertainties were calculated using radius errors from \cite{berger_2020}, and derived \logg errors were calculated using a signal-to-noise metric. Outliers are indicated by outlier flag equal to 1. An inferred \logg of -99 dex corresponds to an unsuccessful inference. An inferred mass of -999 \solmass is set for outliers with inaccurate inferred \logg values. The full table in machine-readable format can be found online.}
\end{deluxetable*}

% 892738 & 11.73 & 4534 & $22.63^{+2.15}_{-1.92}$ & $1.79^{+0.02}_{-0.02}$ & 1.67 & $1.15^{+0.21}_{-0.21}$ & $0.87^{+0.02}_{-0.02}$ & 0\\
%     892760 & 13.23 & 5188 & $9.63^{+0.81}_{-0.74}$ & $2.39^{+0.01}_{-0.01}$ & 2.36 & $1.00^{+0.14}_{-0.14}$ & $0.77^{+0.02}_{-0.02}$ & 0\\
%     893214 & 12.58 & 4728 & $11.30^{+0.83}_{-0.75}$ & $2.53^{+0.01}_{-0.01}$ & 2.54 & $1.72^{+0.10}_{-0.10}$ & $1.61^{+0.03}_{-0.03}$ & 0\\
%     893233 & 11.44 & 4207 & $28.47^{+2.91}_{-2.57}$ & $1.66^{+0.01}_{-0.01}$ & 1.80 & $0.86^{+0.09}_{-0.09}$ & $1.87^{+0.04}_{-0.04}$ & 0\\
%     1026084 & 12.14 & 5072 & $12.79^{+1.17}_{-1.05}$ & $2.58^{+0.01}_{-0.01}$ & 2.56 & $1.71^{+0.18}_{-0.18}$ & $2.18^{+0.04}_{-0.04}$ & 0\\
%     1026309 & 10.60 & 4514 & $21.86^{+1.19}_{-1.11}$ & $2.06^{+0.02}_{-0.02}$ & 2.05 & $2.71^{+0.42}_{-0.42}$ & $1.98^{+0.04}_{-0.04}$ & 0\\
%     1026326 & 13.26 & 5123 & $7.11^{+0.57}_{-0.52}$ & $2.90^{+0.01}_{-0.01}$ & 2.90 & $1.32^{+0.08}_{-0.08}$ & $1.47^{+0.04}_{-0.04}$ & 0\\
%     1026452 & 12.94 & 5089 & $11.99^{+1.26}_{-1.11}$ & $2.46^{+0.01}_{-0.01}$ & 2.47 & $1.55^{+0.17}_{-0.17}$ & $1.55^{+0.04}_{-0.03}$ & 0\\
%     1027110 & 12.10 & 4190 & $26.06^{+2.11}_{-1.89}$ & $1.73^{+0.01}_{-0.01}$ & 1.71 & $1.15^{+0.13}_{-0.13}$ & $1.26^{+0.03}_{-0.03}$ & 0\\
%     1027337 & 12.11 & 4671 & $8.44^{+0.45}_{-0.42}$ & $2.79^{+0.01}_{-0.01}$ & 2.80 & $1.45^{+0.08}_{-0.08}$ & $1.63^{+0.03}_{-0.03}$ & 0 \\

\begin{deluxetable*}{cccccccccc}[t!]
\tabletypesize{\footnotesize}
    \tablecolumns{10}
    \tablewidth{0pt}
    \tablecaption{\label{tb:pande_catalogue} Stellar properties and output parameters for stars in the \gaia sample including derived $\log g$ and mass.}
    \tablehead{\colhead{KIC ID} & \colhead{K$_{\text{p}}$} & \colhead{T$_{\text{eff}}$ [K]} &  \colhead{Radius [$\text{R}_\odot$]} & \colhead{$\log g$ [dex]} & \colhead{$\log g$ Inferred [dex]} & \colhead{True Mass [$\text{M}_\odot$]} & \colhead{Inferred Mass [$\text{M}_\odot$]} & \colhead{SNR} & \colhead{Outlier Flag} }
    \startdata
    1025494 & 11.82 & 5710 & $1.48^{+0.03}_{-0.03}$ & $4.12^{+0.05}_{-0.04}$ & $3.96^{+0.02}_{-0.02}$ & $1.05^{+0.10}_{-0.07}$ & $0.72^{+0.04}_{-0.04}$ & 0.81 & 0\\
    1026475 & 11.87 & 6397 & $1.52^{+0.03}_{-0.03}$ & $4.15^{+0.03}_{-0.04}$ & $4.07^{+0.02}_{-0.02}$ & $1.21^{+0.05}_{-0.08}$ & $0.98^{+0.06}_{-0.06}$ & 0.76 & 0\\
    1026669 & 12.30 & 5924 & $1.14^{+0.02}_{-0.02}$ & $4.28^{+0.04}_{-0.04}$ & $4.13^{+0.02}_{-0.02}$ & $0.91^{+0.07}_{-0.06}$ & $0.64^{+0.04}_{-0.04}$ & 0.75 & 0\\
    1026911 & 12.42 & 6459 & $1.46^{+0.04}_{-0.04}$ & $4.20^{+0.03}_{-0.03}$ & $4.13^{+0.02}_{-0.02}$ & $1.26^{+0.06}_{-0.05}$ & $1.04^{+0.07}_{-0.07}$ & 0.76 & 0\\
    1027030 & 12.34 & 6062 & $1.70^{+0.04}_{-0.04}$ & $4.08^{+0.03}_{-0.04}$ & $3.99^{+0.02}_{-0.02}$ & $1.27^{+0.06}_{-0.10}$ & $1.01^{+0.06}_{-0.07}$ & 0.75 & 0\\
    1163211 & 12.80 & 6080 & $2.23^{+0.07}_{-0.07}$ & $3.85^{+0.06}_{-0.04}$ & $3.95^{+0.02}_{-0.02}$ & $1.30^{+0.16}_{-0.06}$ & $1.63^{+0.13}_{-0.12}$ & 0.77 & 0\\
    1163333 & 13.21 & 6040 & $1.15^{+0.02}_{-0.02}$ & $4.31^{+0.04}_{-0.04}$ & $4.27^{+0.02}_{-0.02}$ & $0.99^{+0.07}_{-0.07}$ & $0.89^{+0.06}_{-0.06}$ & 0.72 & 0\\
    1292493 & 12.56 & 5828 & $1.10^{+0.02}_{-0.02}$ & $4.34^{+0.04}_{-0.04}$ & $4.29^{+0.02}_{-0.02}$ & $0.97^{+0.07}_{-0.07}$ & $0.87^{+0.06}_{-0.05}$ & 0.75 & 0\\
    1292498 & 11.93 & 5734 & $1.15^{+0.02}_{-0.02}$ & $4.29^{+0.04}_{-0.04}$ & $4.16^{+0.02}_{-0.02}$ & $0.94^{+0.07}_{-0.06}$ & $0.70^{+0.04}_{-0.04}$ & 0.78 & 0\\
    1294382 & 12.44 & 6821 & $1.75^{+0.05}_{-0.05}$ & $4.11^{+0.03}_{-0.03}$ & $4.03^{+0.02}_{-0.02}$ & $1.45^{+0.08}_{-0.06}$ & $1.19^{+0.09}_{-0.09}$ & 0.74 & 0\\
    \enddata
    \tablecomments{Stellar parameters such as effective temperature, radius, surface gravity and mass (including uncertainties) were taken from \cite{berger_2020}. Derived mass uncertainties were calculated using radius errors from \cite{berger_2020}, and derived $\log g$ errors were calculated using a signal-to-noise metric. Outliers are indicated by outlier flag equal to 1. An inferred \logg of -99 dex corresponds to an unsuccessful inference. An inferred mass of -999 \solmass is set for outliers with inaccurate inferred \logg values. The full table in machine-readable format can be found online.}
\end{deluxetable*}

\bibliography{references}
\bibliographystyle{aasjournal}

\end{document}